\documentclass[iop, numberedappendix]{emulateapj} 

\usepackage{graphicx}
\usepackage{pslatex}
\usepackage{amsmath, amsthm, amssymb}
\usepackage{natbib}
\usepackage{epsfig}
\usepackage{subfigure}
\usepackage{upgreek}
\usepackage{accents}
\usepackage{hyperref}

\shorttitle{Model Independent Constraints}
\shortauthors{Mirocha et al.}

\begin{document}

%%% Custom definitions
\newcommand*{\dt}[1]{%
  \accentset{\mbox{\large\bfseries .}}{#1}}
\newcommand*{\ddt}[1]{%
  \accentset{\mbox{\large\bfseries .\hspace{-0.25ex}.}}{#1}}

% Physical constants
\newcommand{\kB}{k_{\text{B}}}

% Cosmology
\newcommand{\Omnow}{\Omega_{\text{m},0}}
\newcommand{\Obnow}{\Omega_{\text{b},0}}
\newcommand{\OLnow}{\Omega_{\Lambda,0}}
\newcommand{\Om}{\Omega_{\text{m}}}
\newcommand{\Ob}{\Omega_{\text{b}}}
\newcommand{\OL}{\Omega_{\Lambda}}
\newcommand{\Hnow}{H_0}
\newcommand{\Hofz}{H(z)}

% Various temperatures
\newcommand{\Tcmb}{T_{\gamma}}
\newcommand{\Tcmbnow}{T_{{\gamma},0}}
\newcommand{\Tast}{T_{\ast}}
\newcommand{\zdec}{z_{\text{dec}}}
\newcommand{\TKdec}{T_{\text{K},\text{dec}}}
\newcommand{\TS}{T_{\text{S}}}
\newcommand{\TK}{T_{\text{K}}}
\newcommand{\Tstar}{T_{\star}}

% Shortcuts for common subscripts
\newcommand{\tot}{\text{tot}}
\newcommand{\internal}{\text{int}}
\newcommand{\ioniz}{\text{ion}}
\newcommand{\rec}{\text{rec}}
\newcommand{\recA}{\text{rec,A}}
\newcommand{\recB}{\text{rec,B}}
\newcommand{\igm}{\text{igm}}
\newcommand{\heat}{\text{heat}}

% Lyman-alpha stuff
\newcommand{\Lya}{\text{Ly-}\alpha}
\newcommand{\Lyn}{\text{Ly-}n}
\newcommand{\Ly}{\text{Ly-}}
\newcommand{\LyC}{\text{LyC}}
\newcommand{\nmax}{n_{\text{max}}}
\newcommand{\frec}{f_{\text{rec}}}
\newcommand{\frecn}{f_{\text{rec}}^{(n)}}
\newcommand{\frecbar}{\overline{f}_{\text{rec}}}
\newcommand{\nuLya}{\nu_{\alpha}}
\newcommand{\nuLL}{\nu_{\text{LL}}}

% Densities
\newcommand{\nH}{n_{\text{H}}}
\newcommand{\nHe}{n_{\text{He}}}
\newcommand{\nbar}{\bar{n}^0}
\newcommand{\nHbar}{\bar{n}_{\text{H}}^0}
\newcommand{\nHebar}{\bar{n}_{\text{He}}^0}
\newcommand{\nbbar}{\bar{n}_{\text{b}}^0}
\newcommand{\rhobbar}{\bar{rho}_{\text{b}}^0}

% Emissivities and such
\newcommand{\Jhat}{\widehat{J}_{\alpha}}
\newcommand{\eheat}{\upepsilon_{\heat}}
\newcommand{\eint}{e_{\internal}}
\newcommand{\eX}{\hat{\upepsilon}_X}
\newcommand{\eion}{\hat{\upepsilon}_{\text{ion}}}
\newcommand{\ealpha}{\hat{\upepsilon}_{\alpha}}
\newcommand{\enu}{\hat{\upepsilon}_{\nu}}
\newcommand{\epshat}{\widehat{\upepsilon}}
\newcommand{\enuprime}{\hat{\upepsilon}_{\nu^{\prime}}}

% Species fractions
\newcommand{\xHI}{x_{\text{H } \textsc{i}}}
\newcommand{\xHII}{x_{\text{H } \textsc{ii}}}

\newcommand{\xibar}{\overline{x}_i}

% Hydrogen and helium ions
\newcommand{\HI}{\text{H} {\textsc{i}}}
\newcommand{\HII}{\text{H} {\textsc{ii}}}
\newcommand{\HeI}{\text{He} {\textsc{i}}}
\newcommand{\HeII}{\text{He} {\textsc{ii}}}
\newcommand{\HeIII}{\text{He} {\textsc{iii}}}

% 21-cm features
\newcommand{\zB}{z_{\text{B}}}
\newcommand{\zC}{z_{\text{C}}}
\newcommand{\zD}{z_{\text{D}}}
\newcommand{\znull}{z_{\text{null}}}
\newcommand{\ztrans}{z_{\text{trans}}}
\newcommand{\zfl}{z_{\ast}}
\newcommand{\zbh}{z_{\bullet}}
\newcommand{\zrei}{z_{\text{rei}}}

% BH stuff
\newcommand{\fduty}{f_{\text{duty}}}
\newcommand{\Cedd}{C_{\text{edd}}}
\newcommand{\tedd}{t_{\text{edd}}}
\newcommand{\fedd}{f_{\text{edd}}}
\newcommand{\Mbh}{M_{\bullet}}
\newcommand{\Mdot}{\dot{M}}
\newcommand{\MdotBH}{\dot{M}_{\bullet}}
\newcommand{\mdot}{\dot{m}}
\newcommand{\Ledd}{L_{\text{edd}}}
\newcommand{\BHeff}{\zeta_{\text{acc}}}
\newcommand{\fX}{f_X}
\newcommand{\fXeff}{f_{X,\mathrm{eff}}}

% Random
\newcommand{\xtot}{x_{\tot}}
\newcommand{\coll}{\text{coll}}
\newcommand{\zprime}{z^{\prime}}
\newcommand{\fstar}{f_{\ast}}
\newcommand{\fstarbh}{\tilde{\fstar}}
\newcommand{\fbh}{f_{\bullet}}
\newcommand{\fcoll}{f_{\text{coll}}}
\newcommand{\fcollprime}{f_{\text{coll}}^{\prime}}
\newcommand{\dfcolldz}{\frac{df_{\text{coll}}}{dz}}
\newcommand{\dfcolldztwo}{\frac{d^2f_{\text{coll}}}{dz^2}}
\newcommand{\dfcolldt}{\frac{df_{\text{coll}}}{dt}}
\newcommand{\dfcolldzprime}{\frac{df_{\text{coll}}^{\prime}}{dz}}
\newcommand{\dfcolldtprime}{\frac{df_{\text{coll}}^{\prime}}{dt}}
\newcommand{\dfcolldzbh}{\frac{d\tilde{f}_{\text{coll}}}{dz}}
\newcommand{\dfcolldtbh}{\frac{d\tilde{f}_{\text{coll}}}{dt}}
\newcommand{\mmin}{m_{\text{min}}}
\newcommand{\rhobh}{\rho_{\bullet}}
\newcommand{\rhobhdot}{\dt{\rho}_{\bullet}}
\newcommand{\rhobhdotacc}{\dt{\rho}_{\bullet, \mathrm{acc}}}
\newcommand{\rhobhdotnew}{\dt{\rho}_{\bullet, \mathrm{new}}}
\newcommand{\rhobhdotejec}{\dt{\rho}_{\bullet, \mathrm{ejec}}}
\newcommand{\rhobhdottot}{\dt{\rho}_{\bullet, \tot}}

\newcommand{\rhostar}{\rho_{\ast}}
\newcommand{\rhostardot}{\dt{\rho}_{\ast}}
\newcommand{\rhostarbhdot}{\dt{\rho}_{\ast\bullet}}
\newcommand{\rhom}{\rho_m}
\newcommand{\rhobbarnow}{\bar{\rho}_b^0}
\newcommand{\rhombarnow}{\bar{\rho}_m^0}
\newcommand{\fstardegen}{f_{\ast \bullet}}
\newcommand{\Nion}{N_{\text{ion}}}
\newcommand{\Nalpha}{N_{\alpha}}
\newcommand{\fesc}{f_{\text{esc}}}
\newcommand{\Msun}{M_{\odot}}
\newcommand{\Tvir}{T_{\text{vir}}}
\newcommand{\Tmin}{T_{\text{min}}}
\newcommand{\Tminprime}{T_{\text{min}}^{\prime}}

\newcommand{\nHbarnow}{\bar{n}_{\text{H}}^0}

\newcommand{\fion}{f_{\text{ion}}}
\newcommand{\nnu}{$n_{\nu}$}
\newcommand{\ncol}{N_i}

\newcommand{\zpeak}{z_{\text{peak}}}

\newcommand{\bol}{\mathrm{bol}}
\newcommand{\fbol}{f_{\bol}}

\newcommand{\fheat}{f^{\text{heat}}}
\newcommand{\fXh}{f_{X,h}}
\newcommand{\fioni}{f_i^{\text{ion}}}
\newcommand{\Lbol}{\mathcal{L}_{\text{bol}}}
\newcommand{\spec}{\mathcal{N}}
\newcommand{\Heat}{\mathcal{H}}
\newcommand{\trec}{$t_{\text{rec}}$}
\newcommand{\Lbox}{L_{\mathrm{box}}}
\newcommand{\dx}{\Delta x}
\newcommand{\dd}{\text{d}}

\newcommand{\drIF}{$\Delta r_{\mathrm{IF}}$}
\newcommand{\dTb}{\delta T_b}
\newcommand{\dTbdot}{\dot{\delta T_b}}
\newcommand{\Nvec}{\mathbf{N}}
\newcommand{\sh}{\mathrm{sh}}

% Units
\newcommand{\sfrd}{\Msun \ \mathrm{yr}^{-1} \ \mathrm{cMpc}^{-3}}
\newcommand{\intensityunitsnumber}{\text{s}^{-1} \ \text{cm}^{-2} \ \mathrm{Hz}^{-1} \ \text{sr}^{-1}}
\newcommand{\intensityunitsenergy}{\text{erg} \ \text{s}^{-1} \ \text{cm}^{-2} \ \mathrm{Hz}^{-1} \ \text{sr}^{-1}}
\newcommand{\coemissivityunitsnumber}{\text{s}^{-1} \ \mathrm{Hz}^{-1} \ \text{cMpc}^{-3}}
\newcommand{\coemissivityunitsenergy}{\text{erg} \ \text{s}^{-1} \ \mathrm{Hz}^{-1} \ \text{cMpc}^{-3}}

\newcommand{\emissivityunitsnumber}{\text{s}^{-1} \ \mathrm{Hz}^{-1} \ \text{Mpc}^{-3}}
\newcommand{\emissivityunitsenergy}{\text{erg} \ \text{s}^{-1} \ \mathrm{Hz}^{-1} \ \text{Mpc}^{-3}}

\newcommand{\Jtwoone}{10^{-21} \ \text{erg} \ \text{s}^{-1} \ \text{cm}^{-2} \ \mathrm{Hz}^{-1} \ \text{sr}^{-1}}

\newcommand{\coheatingdensity}{\text{erg} \ \text{s}^{-1} \ \text{cMpc}^{-3}}
\newcommand{\coenergydensity}{\text{erg} \ \text{cMpc}^{-3}}

\newcommand{\dprime}{\prime\prime}
\newcommand{\zdprime}{z^{\dprime}}

\newcommand{\tpB}{$(30.2, -4.8)$}
\newcommand{\tpC}{$(21.1, -112)$}
\newcommand{\tpD}{$(13.5, 24.5)$}

%%% 
%% Title and Affiliations
%%%
\title{Interpreting the Global 21-cm Signal from High Redshifts. I. \\ Model Independent Constraints}
\author{Jordan Mirocha$^{\dagger}$, Geraint J.A. Harker, Jack O. Burns}
\affil{Center for Astrophysics and Space Astronomy, University of Colorado, Campus Box 389, Boulder, CO 80309}
\affil{The NASA Lunar Science Institute, NASA Ames Research Center, Moffett Field, CA 94035, USA}
\email{$^{\dagger}$jordan.mirocha@colorado.edu}

%%%
%% Abstract
%%%
\begin{abstract}
The sky-averaged (global) 21-cm signal is a powerful probe of the
intergalactic medium (IGM) prior to the completion of reionization. However,
it has so far been unclear that even in the best case scenario, in which the
signal is accurately extracted from the foregrounds, that it will provide more
than crude estimates of when the Universe's first stars and black holes form.
In contrast to previous work, which has focused on predicting the 21-cm
signatures of the first luminous objects, we investigate an arbitrary
realization of the signal, and attempt to translate its features to the
physical properties of the IGM. Within a simplified global framework, the
21-cm signal yields quantitative constraints on the $\Lya$ background
intensity, net heat deposition, ionized fraction, and their time derivatives,
without invoking models for the astrophysical sources themselves. The 21-cm
absorption signal is most easily interpreted, setting strong limits on the
heating rate density of the Universe with a measurement of its redshift alone,
independent of the ionization history or details of the $\Lya$ background
evolution. In a companion paper we extend these results, focusing on the
confidence with which one can infer source emissivities from IGM properties.
\end{abstract}
\subjectheadings{dark ages, reionization, first stars, diffuse radiation}

%%%
%% Introduction
%%%
\section{INTRODUCTION} \label{sec:Introduction}
Nearly all of our knowledge about the early Universe comes from the observable
signatures of two phase transitions: the Cosmic Microwave Background (CMB), a
byproduct of cosmological recombination at $z \sim$ 1100 \citep{Spergel2003, Komatsu2011}, and Gunn-Peterson troughs in the spectra of high-$z$ quasars
\citep{Gunn1965}, a sign that cosmological reionization is complete by $z
\gtrsim 6$. The intervening $\sim$Gyr, in which the first stars, black holes,
and galaxies form, is very poorly understood.

Observations with \textit{The Hubble Space Telescope} (HST) have begun to
directly constrain galaxies well into the Epoch of Reionization (EoR) at
redshifts possibly as high as $z \sim 10$ \citep[e.g.][]{Oesch2010, Bouwens2011,
Oesch2012,Zheng2012,Coe2013,Ellis2013}, and upcoming facilities such as \textit{The James Webb Space
Telescope} (JWST) promise to extend this view even further, likely to $z
\gtrsim 10-15$ \citep[e.g.][]{Johnson2009,Zackrisson2012}. However, directly
observing luminous sources at high-$z$ is not equivalent to constraining their
impact on the intergalactic medium (IGM) \citep{Pritchard2007}, be it in the
form of ionization, heating, or more subtle radiative processes (e.g. the
Wouthuysen-Field effect). The most promising probe of the IGM in the
pre-reionization epoch is the redshifted 21-cm signal from neutral hydrogen.
Its evolution over cosmic time encodes the history of heating, ionization, and
$\Lya$ emission, meaning in principle it is a probe of the background
intensity at photon energies ranging from the $\Lya$ resonance to hard X-rays
\citep[for a recent review, see][]{FurlanettoOhBriggs2006}.

At stake in the quest to observe the Universe in its infancy is an
understanding of galaxy formation, which currently rests upon a theoretically
reasonable but virtually unconstrained foundation. The first stars are
expected to be very massive \citep[$M \gtrsim 100\ \Msun$; e.g.][]{Haiman1996,
Tegmark1997, Bromm1999, Abel2002b} resulting in surface temperatures of order
$10^5$ K \citep{Tumlinson2000, Bromm2001, Schaerer2002}, though evidence for
such objects is currently limited to abundance patterns in low-mass stars in
the Milky Way \citep[e.g.][]{Salvadori2007,Rollinde2009}. Whether or not such
massive stars ever form is a vital piece of the galaxy formation puzzle, as
their brief existence is expected to dramatically alter the physical
conditions for subsequent star formation: first, through an intense soft-UV
radiation field, which photo-ionizes (dissociates) atomic (molecular)
hydrogen, and presumably via metal enrichment and thermal feedback following a
supernovae explosion \citep[see review by][]{Bromm2009}.

Even if the first stars are $\sim 100 \ \Msun$ and leave behind remnant black
holes of comparable mass, it is difficult to reconcile the existence of $z
\gtrsim 6-7$ quasars \citep{Fan2006, Mortlock2011}, whose luminosities imply
accretion onto super-massive black holes (SMBHs) with masses $\Mbh \gtrsim
10^9 \ \Msun$, with models of growth via Eddington-limited accretion. The
difficulty of growing SMBHs from modest seeds has inspired direct-collapse
models \citep{Begelman2006, Begelman2008}, which predict the formation of BHs
with $\Mbh \gtrsim 10^3 \ \Msun$ in massive, atomic-cooling dark matter halos
via dynamical instabilities. These models alleviate the requirement of
continual Eddington-limited accretion throughout the reionization epoch, but
remain unconstrained.

\textit{JWST} may be able to detect clusters of PopIII stars at $2 \lesssim z
\lesssim 7$ \citep{Johnson2010}, PopIII galaxies and quasistars at $z \sim
10-15$ \citep{Zackrisson2011,Johnson2012}, and PopIII supernovae at $z \sim
15-20$ \citep{Whalen2013SNIIn,Whalen2013PISN}, depending on their masses,
emission properties, etc. However, the prospects for constraining the
\textit{first} generations of stars and black holes via direct detection,
which likely form at higher redshifts, are bleak. The prospects for
constraining the first stars and black holes \textit{indirectly}, however, are
encouraging at low radio frequencies, regardless of their detailed properties.

While the long term goal is to map the 21-cm fluctuations from the ground
\citep[a task on the horizon at $z \lesssim 10$; e.g. LOFAR, MWA, PAPER, GMRT,
SKA;
][]{Harker2010,vanHaarlem2013,Bowman2013,Parsons2010,Paciga2013,Carilli2004,Mellema2013})
or space \citep[e.g. the Lunar Radio Array (LRA);][]{Jester2009} using large
interferometers, in the near term, the entire $10 \lesssim z \lesssim 40$
window is likely to be accessible only to all-sky 21-cm experiments. Several
challenges remain, however, from both observational and theoretical
perspectives. The Earth is a sub-optimal platform for observations at the
relevant frequencies ($\nu \lesssim 200 \ \mathrm{MHz}$) due to
radio-frequency interference (RFI) and ionospheric variability
\citep{Vedantham2013}, making the lunar farside a particularly appealing
destination for future observatories \citep[e.g. the LRA, \textit{The Dark
Ages Radio Explorer} (DARE);][]{Burns2012}. Some foregrounds cannot be escaped
even from the lunar farside (e.g. synchrotron emission from our own galaxy),
and must be removed in post-processing using sophisticated fitting algorithms
\citep[e.g.][]{Harker2012,Liu2013}. To date, ground based 21-cm efforts have
largely focused on the end of the EoR ($100 \lesssim [\nu / \mathrm{MHz}]
\lesssim 200$), including lower limits on the duration of reionization
\citep[via the single-element EDGES instrument;][]{Bowman2010}, and
constraints on the thermal and ionization history with single dish telescopes
and multi-element interferometers \citep[e.g.][]{Paciga2013,Parsons2013}.
Extending this view to ``cosmic dawn'' requires observations below 100 MHz, a frequency range most easily explored from the radio-quiet,
ionosphere-free\footnote{The Moon is not truly devoid of an ionosphere -- its
atmosphere is characterized as a surface-bounded exosphere, whose constituents
are primarily metal ions liberated by interactions with energetic particles
and radiation from the Sun \citep[e.g.][]{Stern1999}. However, it is tenuous
enough to be neglected at frequencies $\nu \gtrsim 1 \ \mathrm{MHz}$. },
lunar farside.

Even if the astrophysical signal is extracted from the foregrounds
perfectly, it is not clear that one could glean more than gross
estimates of the timing of first star and black hole formation. While simply
knowing the redshift at which the first stars and black holes form would be an
enormous achievement, ultimately it is their properties that are of interest.
Were the Universe's first stars very massive? Did all SMBHs in the local
Universe form via direct collapse at high-$z$? Could the global 21-cm signal
alone rule out models for the formation of the first stars and black holes?
What if independent measurements from JWST and/or other facilities were
available?

Motivated by such questions we turn our attention to the final stage of any
21-cm pipeline: interpreting the measurement. Rather than formulating
astrophysical models and studying 21-cm realizations that result, we focus on
an \textit{arbitrary} realization of the signal, and attempt to recover the
properties of the Universe in which it was observed. We defer a
detailed discussion of how these properties of the Universe (e.g. the
temperature, ionized fraction, etc.) relate to astrophysical sources to Paper
II (Mirocha et al., in prep).

The outline of this paper is as follows. In Section 2, we introduce the physical processes that give rise to the 21-cm signal. In Section 3, we step through the three expected astrophysical features of the signal, focusing on how observational measures translate to physical properties of the Universe. Discussion and conclusions are presented in Sections 4 and 5, respectively.

We adopt a cosmology with $\Omnow=0.272$, $\Obnow=0.044$, $\OLnow=0.728$, and $\Hnow=70.2 \ \mathrm{km} \ \mathrm{s}^{-1} \ \mathrm{Mpc}^{-3}$ throughout.

%%% 
%% Formalism
%%%
\section{Formalism} \label{sec:Formalism}

% 21-cm
\subsection{Magnitude of the 21-cm Signal}
The 21-cm transition results from hyperfine splitting in the $1\mathrm{S}$
ground state of the hydrogen atom when the magnetic moments of the proton and
electron flip between aligned (triplet state) and anti-aligned (singlet
state). The HI brightness temperature depends sensitively on the ``spin
temperature,'' $\TS$, a 21-cm specific excitation temperature
which characterizes the number of hydrogen atoms in the triplet and singlet
states, $(n_1/n_0) = (g_1/g_0) \exp(-\Tstar/\TS)$, where $g_1$ and $g_0$ are
the degeneracies of the triplet and singlet hyperfine states, respectively,
and $\Tstar = 0.068$ K is the temperature corresponding to the energy
difference between hyperfine levels.

The redshift evolution of the 21-cm signal, $\dTb(z)$, as measured relative to
the CMB, also depends on the mean hydrogen ionized fraction, $\xibar$, and in
general on the baryon over-density and proper motions along the line of sight,
though the last two effects should be negligible for studies of the all-sky
spectrum, leaving \citep[e.g][]{FurlanettoOhBriggs2006},
\begin{equation}
    \dTb \simeq 27 (1 - \xibar) \left(\frac{\Obnow h^2}{0.023} \right) \left(\frac{0.15}{\Omnow h^2} \frac{1 + z}{10} \right)^{1/2} \left(1 - \frac{\Tcmb}{T_{\mathrm{S}}} \right) , \label{eq:dTb}
\end{equation}
where $h$ is the Hubble parameter today in units of $100 \ \mathrm{km} \
\mathrm{s}^{-1} \ \mathrm{Mpc}^{-1}$, and $\Obnow$ and $\Omnow$ are the
fractional contributions of baryons and matter to the critical energy density,
respectively.

Whether the signal is seen in emission or absorption against the CMB depends entirely on the spin temperature, which is determined by the strength of collisional coupling and presence of background radiation fields,
\begin{equation}
    T_S^{-1} \approx \frac{T_{\gamma}^{-1} + x_c T_K^{-1} + x_{\alpha} T_{\alpha}^{-1}}{1 + x_c + x_{\alpha}} \label{eq:SpinTemperature}
\end{equation}
where $\Tcmb = \Tcmbnow(1+z)$ is the CMB temperature, $\TK$ is the kinetic temperature, and $T_{\alpha} \approx \TK$ is the UV color temperature. 

In general, the collisional coupling is a sum over collision-partners,
\begin{equation}
    x_c = \sum_i \frac{n_i \kappa_{10}^i}{A_{10}} \frac{\Tast}{\Tcmb} \label{eq:xc}
\end{equation}
where $n_i$ is the number density of species $i$, and $\kappa_{10}^i =
\kappa_{10}^i(\TK)$ is the rate coefficient for spin de-excitation via
collisions with species $i$. In a neutral gas, collisional coupling is
dominated by hydrogen-hydrogen collisions \citep{Allison1969, Zygelman2005,
Sigurdson2006}, though hydrogen-electron collisions can become important as
the ionized fraction and temperature grow \citep{FurlanettoFurlanetto2007a}.
We neglect collisional coupling due to all other
species\footnote{\citet{FurlanettoFurlanetto2007b} investigated the effects of
hydrogen-proton collisions on $\TS$ and found that they could account for up
to $\sim 2$\% of the collisional coupling at $z \approx 20$, and would
dominate the coupling at $z\approx 10$ in the absence of heat sources.
However, an early $\Lya$ background is expected to couple $\TS\rightarrow \TK$
prior to $z = 20$, and heating is expected prior $z = 10$, so protons are
generally neglected in 21-cm calculations. Collisions with neutral helium
atoms in the triplet state could also induce spin-exchange \citep{Hirata2007},
though the cold high-$z$ IGM lacks the energy required to excite atoms to the
triplet state. We also neglect hydrogen-deuterium collisions, whose rarity
prevents any real effect on $\TS$, even though $\kappa_{10}^{\mathrm{HD}} >
\kappa_{10}^{\mathrm{HH}}$ at low temperatures \citep{Sigurdson2006}. Lastly,
we neglect velocity-dependent effects \citep{Hirata2007}, which introduces an
uncertainty of up to a few \% in the mean signal.}.

The remaining coupling coefficient, $x_{\alpha}$, characterizes the strength of Wouthuysen-Field coupling \citep{Wouthuysen1952,Field1958},
\begin{equation}
    x_{\alpha} = \frac{S_{\alpha}}{1+z} \frac{\Jhat}{\overline{J}_{\alpha}} \label{eq:Jalpha}
\end{equation}
where
\begin{equation}
    \overline{J}_{\alpha} \equiv \frac{16\pi^2 \Tstar e^2 f_{\alpha}}{27 A_{10} \Tcmbnow m_e c} . 
\end{equation}    
$\Jhat$ is the angle-averaged intensity of $\Lya$ photons in units of
$\intensityunitsnumber$, $S_{\alpha}$ is a correction factor that accounts for
variations in the background intensity near line-center \citep{Chen2004,
FurlanettoPritchard2006, Hirata2006}, $m_e$ and $e$ are the electron mass and
charge, respectively, $f_{\alpha}$ is the $\Lya$ oscillator strength, and
$A_{10}$ is the Einstein A coefficient for the 21-cm transition.

% Inflection points
\subsection{Slope of the 21-cm Signal}
\begin{figure}[htbp]
\begin{center}
\includegraphics[width=0.48\textwidth]{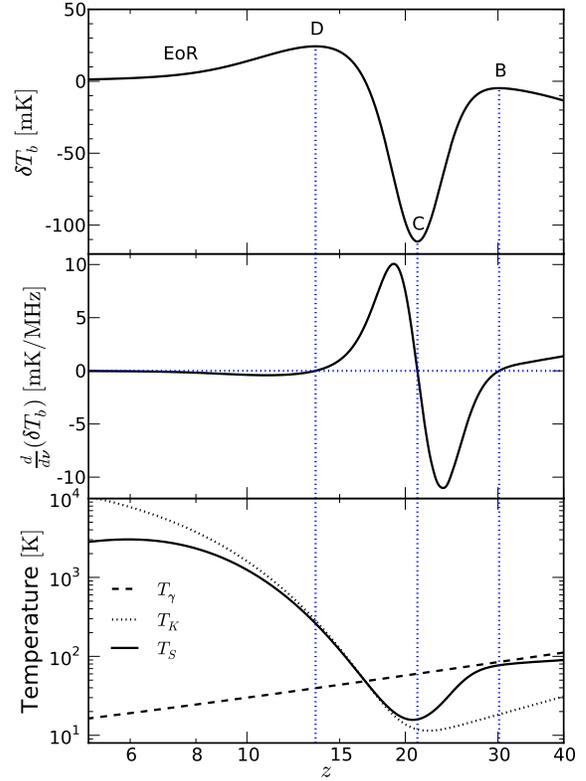}
\caption{An example global 21-cm spectrum (top), its derivative (middle), and corresponding thermal evolution (bottom) for a model in which reionization is driven by PopII stars, and the X-ray emissivity of the Universe is dominated by high-mass X-ray binaries.}
\label{fig:global_signal}
\end{center}
\end{figure}

Models for the global 21-cm signal generally result in a curve with five
extrema\footnote{We neglect the first and last features of the
signal in this paper. The lowest redshift feature marks the end of
reionization, and while its frequency derivative is zero, so is its amplitude,
making its precise location difficult to pinpoint. The highest redshift
feature is neglected because it is well understood theoretically, and should
occur well before the formation of the first luminous objects \citep[though
exotic physics such as dark-matter annihilation could complicate this,
e.g.][]{FurlanettoOh2006}.}, three of which are labeled in Figure
\ref{fig:global_signal}, roughly corresponding to the formation of the first
stars (B), black holes (C), and beginning of the EoR (D). Due to the
presence of strong \citep[but spectrally smooth in principle; see][]{Petrovic2011} foregrounds, the ``turning points'' are likely the
only pieces of the signal that can be reliably extracted
\citep[e.g.][]{Pritchard2010a, Harker2012}. Our primary goal in
\S\ref{sec:CriticalPoints} will be to determine the quantitative physical
meaning of each feature in turn.

In preparation, we differentiate Equation \ref{eq:dTb},
\begin{align}
    \frac{d}{d\nu} & \bigg[\dTb \bigg] \simeq 0.1 \left(\frac{1 - \xibar}{0.5}\right) \left(\frac{1 + z}{10}\right)^{3/2} \left\{\left(\frac{\Tcmb}{\TS} \right) \left[1 +  \frac{3}{2}\frac{d\log T_S}{d\log t}\right] \right. \nonumber \\
    & \left. - \frac{1}{2(1 - \xibar)} \left(1 - \frac{\Tcmb}{\TS} \right) \left[1 - \xibar \left(1 - 3 \frac{d\log \xibar}{d\log t}\right) \right] \right\} \mathrm{mK} \ \mathrm{MHz}^{-1}
 \label{eq:dTbdnu} .
\end{align} 
making it clear that at an extremum, the following condition must be satisfied:
\begin{equation}
    \frac{d\log T_S}{d\log t} = \frac{1}{3(1 - \xibar)} \left(\frac{\TS}{\Tcmb} - 1\right)\left[1 - \xibar\left(1 - 3 \frac{d\log \xibar}{d\log t}\right)\right] - \frac{2}{3} \label{eq:TurningPoint}
\end{equation}
We can obtain a second independent equation for the spin-temperature rate of change by differentiating Equation \ref{eq:SpinTemperature},
\begin{align}
    \frac{d\log T_S}{d\log t} & = \left[1 + x_{\tot} \left(\frac{\Tcmb}{\TK}\right) \right]^{-1} \left\{\frac{x_{\tot}}{(1 + x_{\tot})} \frac{d\log x_{\tot}}{d\log t} \left[1 - \left(\frac{\Tcmb}{\TK}\right)\right] \right. \nonumber \\
    & \left. + x_{\tot} \frac{d\log T_K}{d\log t} \left(\frac{\Tcmb}{T_K}\right) - \frac{2}{3} \right\} \label{eq:dlogTs} .
\end{align}
where $\xtot = x_c + x_{\alpha}$, such that
\begin{equation}
    \frac{d\log x_{\tot}}{d\log t} = x_{\tot}^{-1} \left[\sum_i x_c^i \frac{d\log x_c^i}{d\log t}   + x_{\alpha}\frac{d\log x_{\alpha}}{d\log t} \right] \label{eq:dlogxtot} .
\end{equation} 
Expanding out the derivatives of the coupling terms, we have
\begin{equation}
    \frac{d\log x_{\alpha}}{d\log t} = \frac{d\log \Jhat}{d\log t} + \frac{d\log S_{\alpha}}{d\log \TK} \frac{d\log \TK}{d\log t} + \frac{2}{3} \label{eq:dlogxa} 
\end{equation}
and
\begin{equation}
   \frac{d\log x_c^i}{d\log t} = \frac{d\log \kappa_{10}^i}{d\log \TK} \frac{d\log \TK}{d\log t} \pm \frac{d\log x_e}{d\log t} - \frac{4}{3} \label{eq:dlogxc}
\end{equation}
where the second to last term is positive for H-H collisions and negative for
H-$e^-$ collisions. 

As in \citet{Furlanetto2006} and \citet{Pritchard2007}, we adopt a two-zone
model in which the volume filling fraction of HII regions, $x_i$, is treated
separately from the ionization in the bulk IGM, parameterized by $x_e$. The
mean ionized fraction is then $\xibar = x_i + (1 - x_i) x_e$. This treatment
is motivated\footnote{Our motivation for the logarithmic derivative convention
is primarily compactness, though the non-dimensionalization of derivatives is
convenient for comparing the rate at which disparate quantities evolve. For
reference, the logarithmic derivative of a generic function of redshift with
respect to time, $d\log w/d\log t = b$, implies $w(z) \propto (1 + z)^{-3b/2}$
under the high-$z$ approximation, $H(z) \approx H_0 \Omnow^{1/2} (1 +
z)^{3/2}$, which is accurate to better than $\sim 0.5$\% for all $z > 6$. For
example, the CMB cools as $d\log \Tcmb/d\log t = -2/3$.} by the fact that
$\dTb=0$ in HII regions, thus eliminating the need for a detailed treatment of
the temperature and ionization evolution, but beyond HII regions, the gas is
warm and only partially ionized (at least at early times) so we must track
both the kinetic temperature and electron density in order to compute the spin
temperature.

%%
% Critical Points in the 21-cm History
%%
\section{CRITICAL POINTS IN THE 21-CM HISTORY} \label{sec:CriticalPoints}
From the equations of \S\ref{sec:Formalism}, it is clear that in general,
turning points in the 21-cm signal probe a set of eight quantities,
$\boldsymbol{\theta} = \{x_i, x_e, \TK, \Jhat, x_i^{\prime}, x_e^{\prime},
\TK^{\prime}, \Jhat^{\prime}\}$, where primes represent logarithmic time
derivatives. Given a perfect measurement of the redshift and brightness
temperature, $(z, \dTb)$, at a turning point, the system is severely
underdetermined with two equations (Eqs. \ref{eq:dTb} and
\ref{eq:TurningPoint}) and eight unknowns. Without independent measurements of
the thermal and/or ionization history and/or $\Lya$ background intensity, no
single element of $\boldsymbol{\theta}$ can be constrained unless one or more
assumptions are made to reduce the dimensionality of the problem.

The most reasonable assumptions at our disposal are:
\begin{enumerate}
    \item The volume filling factor of HII regions, $x_i$, and the ionized fraction in the bulk IGM, $x_e$, are both negligible, as are their time derivatives, such that $\xibar = d\log \xibar/d\log t = 0$.
    \item There are no heat sources, such that the Universe's temperature is governed by pure adiabatic cooling after decoupling at $\zdec \simeq 150$ \citep{Peebles1993}, i.e. $d\log \TK / d\log t = -4/3$.
    \item $\Lya$ coupling is strong, i.e. $x_{\alpha} \gtrsim 1$, such that $\TS \rightarrow \TK$, and the dependencies on $\Jhat$ no longer need be considered.
\end{enumerate}      
These assumptions are expected to be valid at $z \gtrsim$ 10, $z \gtrsim 20$,
and $z \lesssim 10$, respectively, according to typical models
\citep[e.g.][]{Furlanetto2006, Pritchard2010a}. But, since it may be
impossible to verify their validity from the 21-cm signal alone, we will take
care in the following sections to state explicitly how each assumption affects
inferred values of $\boldsymbol{\theta}$. We will now examine each feature of
the signal in turn.

%%
% Turning Point B
%%
\subsection{Turning Point B: End of the Dark Ages} \label{sec:B}
Prior to the formation of the first stars, the Universe is neutral to a part
in $\sim 10^4$ \citep[e.g. \texttt{RECFAST}, \texttt{HyRec},
\texttt{CosmoRec};][]{Seager1999, Seager2000, AliHamoud2010, Chluba2011}, such that a measurement of $\dTb$ probes $\TS$ directly via Equation
\ref{eq:dTb}, 
\begin{equation}
    \TS \leq \Tcmb \left[1 - \frac{\dTb}{9 \ \mathrm{mK}} (1 + z)^{-1/2} \right]^{-1}  \label{eq:TurningPointNeutral_TS}
\end{equation}
where the $\leq$ symbol accounts for the possibility that $\xibar > 0$ (a
non-zero ionized fraction always acts to reduce the amplitude of the signal). For the first generation of objects, we can safely assume $\xibar \ll 1$, and
interpret a measurement of the brightness temperature as a proper constraint
on $\TS$ (rather than an upper limit). We will relax this requirement in
\S\ref{sec:C}.

\begin{figure}[htbp]
\begin{center}
\includegraphics[width=0.48\textwidth]{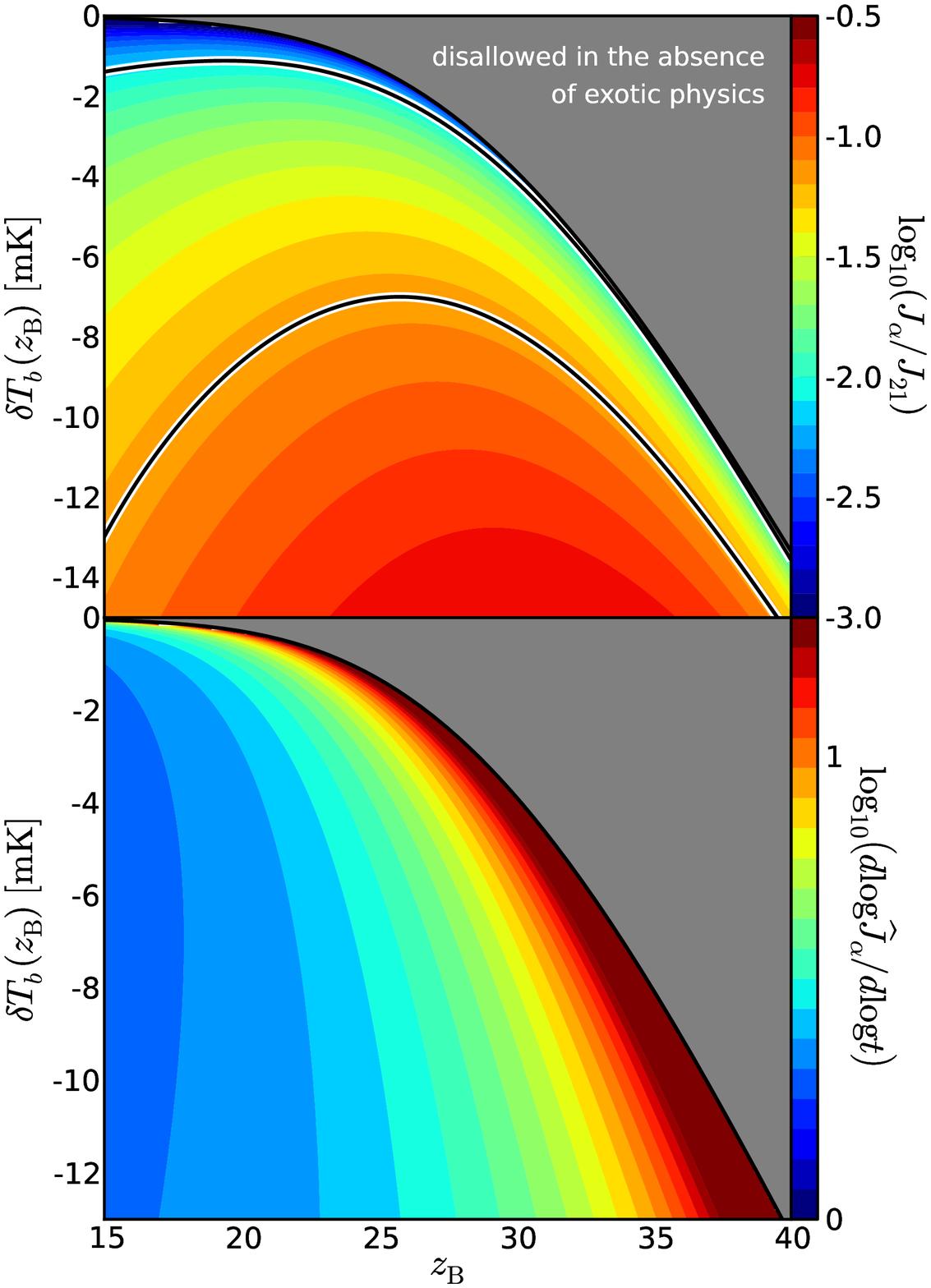}
\caption{Values $J_{\alpha} = h\nuLya \Jhat$ and $d\log J_{\alpha} /
d\log t$ that give rise to turning point B at position $(\zB, \dTb(\zB))$. The
color scale shows the value of $J_{\alpha}$ (top panel, in units of $J_{21} =
\Jtwoone$), and $d\log \Jhat / d\log t$ (bottom panel) required for
turning point B to appear at the corresponding position in the $(\zB,
\dTb(\zB))$ plane, under the assumptions given in Section 3.1. The gray shaded
region is excluded unless heating occurs in the dark ages. For reference, the
highlighted black contours represent $\Lya$ fluxes (assuming a flat spectral
energy distribution at energies between $\Lya$ and the Lyman-limit,
$h\nuLya \leq h\nu \leq h\nu_{\mathrm{LL}}$), corresponding to
Lyman-Werner band fluxes of $J_{\mathrm{LW}} / J_{21} = \{10^{-2}, 10^{-1}\}$
(from top to bottom), which roughly bracket the range of fluxes expected to
induce negative feedback in minihalos at $z \sim 30$ \citep{Haiman2000}.}
\label{fig:tpB}
\end{center}
\end{figure}

If $\TS$ and $\TK$ are both known, Equation \ref{eq:SpinTemperature} yields
the total coupling strength, $\xtot$. But, the contribution from collisional
coupling is known as a function of redshift for a neutral
adiabatically-cooling gas, and can simply be subtracted from $\xtot$ to yield
$x_{\alpha}$, and thus $\Jhat$ (via Eq. \ref{eq:Jalpha}). The top
panel of Figure \ref{fig:tpB} shows lines of constant $\log_{10} (J_{\alpha} /
J_{21})$, where $J_{\alpha} = h\nuLya \Jhat$ and $J_{21} = \Jtwoone$,
given the redshift and brightness temperature of turning point B,
$\dTb(\zB)$. From Equations \ref{eq:TurningPoint} and \ref{eq:dlogTs}, we can
also constrain the rate of change in the background $\Lya$ intensity (Eq.
\ref{eq:dlogxa}), as shown in the bottom panel of Figure \ref{fig:tpB}.

%\footnote{The relationship between $d\log \Jhat / d\log t$ and the co-moving $\Lya$ emissivity, $\ealpha$, could in general be very complex. While this will be addressed in an upcoming paper in detail, for now we simply note for intuition's sake that for a population of monochromatic $\Lya$ emitters (neglecting all radiative transfer effects, i.e. $\Jhat \propto \ealpha$), $d\log \Jhat / d\log t = d\log \ealpha / d\log t - 2$.}

In the event that heating has already begun (rendering $\TK(z)$ unknown),
interpreting turning point B becomes more complicated\footnote{We
deem such a scenario ``exotic'' because it requires heat sources prior to the
formation of the first stars. Heating via dark matter annihilation is one
example of such a heating mechanism \citep{FurlanettoOh2006}.}. Now,
$x_{\alpha}$ will be overestimated, given that a larger (unknown) fraction of
$\xtot$ is due to collisional coupling. Uncertainty in $\TK$ propagates to
$S_{\alpha}$, meaning $x_{\alpha}$ can only be considered to provide an upper
limit on the product $S_{\alpha} \Jhat$, rather than $\Jhat$ alone.
Interpretation of the turning point condition (Eq. \ref{eq:TurningPoint})
becomes similarly complicated if no knowledge of $\TK(z)$ is assumed.

%%
% Turning Point C
%%
\subsection{Turning Point C: Heating Epoch} \label{sec:C}
In the general case where Hubble cooling and heating from astrophysical
sources must both be considered, the temperature evolution can be written as
\begin{equation} 
    \frac{d\log \TK}{d\log t} = \frac{\tau_H}{\tau_X} -
\mathcal{C} \label{eq:ThermalEvolution} 
\end{equation}
where we've defined a characteristic heating timescale $\tau_X^{-1} \equiv
\eheat / \eint$, where $\eint$ is the gas internal energy, $\eheat$ and
$\mathcal{C}$ are the heating and cooling rate densities, respectively, and
$\tau_H^{-1} = 3 \Hofz / 2$ is a Hubble time at redshift $z$ in a
matter-dominated Universe.

\begin{figure}[htbp]
\begin{center}
\includegraphics[width=0.48\textwidth]{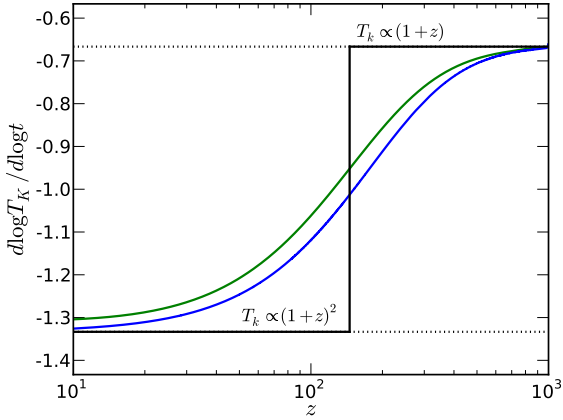}
\caption{Cooling rate of the Universe under different assumptions. The black
line is an approximate analytic solution \citep{Peebles1993}, while
the blue and green lines are numerical solutions. The blue curve considers
cooling via radiative recombination, collisional excitation and ionization,
and the Hubble expansion, and heating via Compton scattering. The green
line is an even more detailed numerical solution obtained with the
\texttt{CosmoRec} code \citep{Chluba2011}, which includes a multi-level atom
treatment and many radiative transfer effects.}
\label{fig:thermal_evolution}
\end{center}
\end{figure}

In a neutral medium, the solution to Equation \ref{eq:ThermalEvolution} for an arbitrary $\eheat$ is 
\begin{equation}
    \TK(z) = \mathcal{C}_1^{-1} \int_{z}^{\infty} \eheat(z^{\prime}) \frac{dt}{dz^{\prime}} dz^{\prime} + \Tcmbnow \frac{(1 + z)^2}{1+\zdec}  \label{eq:TemperatureSolution}
\end{equation}    
where $\mathcal{C}_1 \equiv 3 \nHbar (1 + y) \kB / 2$, $\kB$
is Boltzmann's constant, $\nHbar$ is the hydrogen number density today, $y$ is
the primordial helium abundance (by number), and the second term represents
the adiabatic cooling limit.

To move forward analytically we again adopt the maximal cooling rate,
$\mathcal{C} = 4/3$. Detailed calculations with \texttt{CosmoRec} indicate
that such a cooling rate is not achieved until $z \lesssim 10$ in the absence
of heat sources, which means we \textit{overestimate} the cooling rate, and
thus \textit{underestimate} $\TK$ at all redshifts. This lower bound on the
temperature is verified in Figure \ref{fig:thermal_evolution}, in which we
compare three different solutions for the cooling rate density evolution.

\begin{figure}[htbp]
\begin{center}
\includegraphics[width=0.48\textwidth]{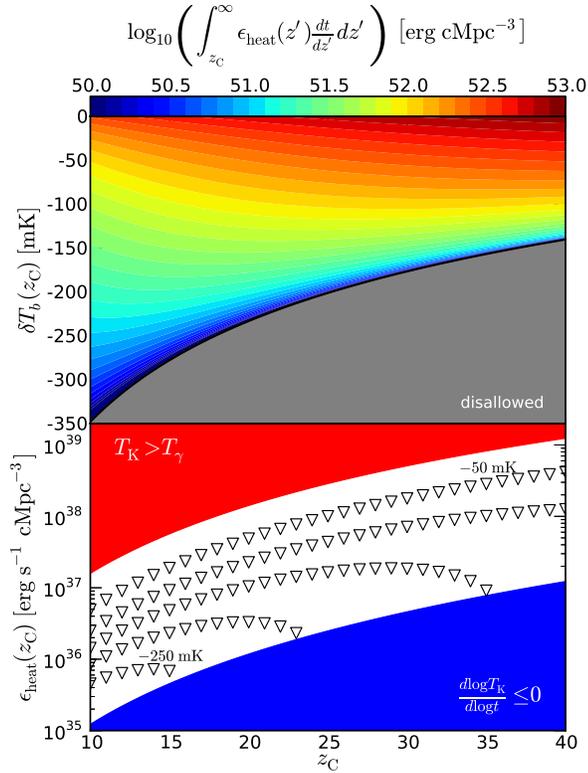}
\caption{\textit{Top:} Constraints on the cumulative energy deposition as a function of the redshift and brightness temperature of turning point C. The gray region is disallowed because it requires cooling to be more rapid than Hubble (adiabatic) cooling. \textit{Bottom:} Constraints on the co-moving heating rate density ($\mathrm{cMpc}^{-3}$ means co-moving $\mathrm{Mpc}^{-3}$) as a function of $\zC$ alone. The blue region includes heating rate densities insufficient to overcome the Hubble cooling, while the red region is inconsistent with the existence of an absorption feature at $\zC$ because such heating rates would instantaneously heat $\TK$ above $\Tcmb$. The triangles, plotted in increments of $50$ mK between $\dTb = \{-250,-50\}$ mK show how a measurement of $\dTb(\zC)$, as opposed to $\zC$ alone, enables more stringent upper limits on the heating rate density.}
\label{fig:tpC_heat_constraints}
\end{center}
\end{figure}

In order for the 21-cm signal to approach emission, the temperature must be
increasing relative to the CMB\footnote{Though see \S\ref{sec:Cion} for an
alternative scenario.}, i.e. $\tau_H/\tau_X > 4/3$, meaning the existence of
turning point C, at redshift $\zC$, alone gives us a lower limit on
$\eheat(\zC)$. Detection of the absorption signal (regardless of its
amplitude) also requires the kinetic temperature to be cooler than the CMB
temperature. If we assume a `burst' of heating, $\eheat \rightarrow \eheat
\delta(z - \zC)$, where $\delta$ is the Dirac Delta function, and require $\TK
< \Tcmb$, we can solve Equation \ref{eq:ThermalEvolution} and obtain an upper
limit on the co-moving heating rate density. The bottom panel of Figure
\ref{fig:tpC_heat_constraints} shows the upper and lower limits on $\eheat$ as
a function of $\zC$ alone.

A stronger upper limit on $\eheat(\zC)$ is within reach, however, if we can
measure the brightness temperature of turning point C accurately. Given that
$\dTb(\zC)$ provides an upper limit on $\TS$ for all values of $\xibar$ (Eq.
\ref{eq:TurningPointNeutral_TS}), and an absorption signal requires $\TK <
\TS < \Tcmb$, we can solve Equation \ref{eq:TemperatureSolution} assuming
$\TK < \TS$, and once again assume a burst of heating to get a
revised upper limit on $\eheat(\zC)$. 

In general, turning point C yields an upper limit (again because we've assumed
$\mathcal{C} = 4/3$) on the \textit{integral} of the heating rate density
(Eq. \ref{eq:TemperatureSolution}), which is seen in the upper panel of Figure
\ref{fig:tpC_heat_constraints}\footnote{We express our results in units of $\mathrm{erg} \ \mathrm{cMpc}^{-3}$ to ease the conversion between $\eheat$ and the X-ray emissivity, $\eX$ (see \S\ref{sec:ExampleHistory}). For reference, $10^{51} \ \mathrm{erg} \ \mathrm{cMpc}^{-3} \simeq 10^{-4} \ \mathrm{eV} \ \mathrm{baryon}^{-1}$.}. This upper limit is independent of the
ionization history, since any ionization reduces the amplitude of $\TS$, thus
lessening the amount of heating required to explain an absorption feature of a
given depth. The only observational constraints available to date are
consistent with X-ray heating of the IGM at $z \gtrsim 8$ \citep{Parsons2013}.

% Zero-crossing
\subsubsection{From Absorption to Emission} \label{sec:trans}
\begin{figure}[htbp]
\begin{center}
\includegraphics[width=0.48\textwidth]{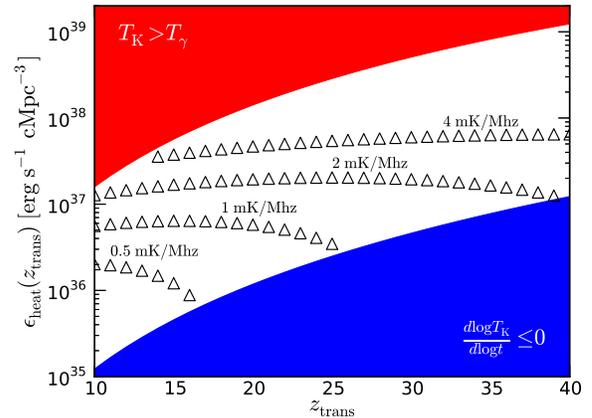}
\caption{Constraints on the co-moving heating rate density (once again $\mathrm{cMpc}^{-3}$ means co-moving $\mathrm{Mpc}^{-3}$) as a function of the absorption-emission transition redshift, $\ztrans$, and the slope of the 21-cm signal at that redshift. As in Figure \ref{fig:tpC_heat_constraints}, the blue region indicates heating rates insufficient to overcome the Hubble cooling, while the red region denotes heating rates that would instantaneously heat $\TK$ above $\Tcmb$. The triangles show how measuring the slope of the signal at $z_{\mathrm{trans}}$ can provide a lower limit on $\eheat$.}
\label{fig:slope_constraints}
\end{center}
\end{figure}

If heating persists, and the Universe is not yet reionized, the 21-cm signal will eventually transition from absorption to emission. At this time, coupling is expected to be strong such that at the precise redshift of the transition, $z_{\mathrm{trans}}$, Equation \ref{eq:dTbdnu} takes special form since $\TS \simeq \TK = \Tcmb$, 
\begin{align}
    \frac{d}{d\nu}\bigg[\dTb \bigg] & \simeq 0.1 \left(\frac{1 - x_i}{0.5}\right) \left(\frac{1 + z_{\mathrm{trans}}}{10}\right)^{3/2} \nonumber \\
    & \times \left[1 +  \frac{3}{2}\frac{d\log \TK}{d\log t}\right] \mathrm{mK} \ \mathrm{MHz}^{-1} . \label{eq:TransitionSlope}
\end{align}
That is, if we can measure the slope at the absorption-emission transition, we
obtain a lower limit on the heating rate density. Our inferred heating rate
density would be exact if $\xibar$ were identically zero, but for $\xibar >
0$, the slope provides a lower limit. This is illustrated in the Figure
\ref{fig:slope_constraints}.

% Could turning point C be a by-product of ionization?
\subsubsection{Could the Absorption Feature be Ionization-Driven?} \label{sec:tpC_ion} \label{sec:Cion}
The absorption feature of the all-sky 21-cm signal is generally expected to
occur when X-rays begin heating the IGM
\citep[e.g.][]{Ricotti2005,Ciardi2010}. However, this feature could also be
produced given sufficient ionization, which similarly acts to drive the signal
toward emission (albeit by reducing the absolute value of $\dTb$ rather than
increasing $\TS$). We now assess whether or not such a scenario could produce
turning point C while remaining consistent with current constraints from
the Thomson optical depth to the CMB \citep[$\tau_e$;][]{Dunkley2009,
Larson2011,Bennett2012}.

We assume that coupling is strong, $\TS \simeq \TK$, and that the Universe
cools adiabatically (i.e. the extreme case where turning point C is
\textit{entirely} due to ionization), so that a measurement of $\dTb$ is a
direct proxy for the ionization fraction (via Eq. \ref{eq:dTb}). If we adopt a
$\mathrm{tanh}$ model of reionization, parameterized by the midpoint of
reionization, $\zrei$, and its duration, $\Delta \zrei$, we can solve Equation
\ref{eq:dTb} at a given $\dTb(\zC)$ for $\xibar(\zC)$. Then, we can determine
the ($\zrei$, $\Delta \zrei$) pair, and thus entire ionization history
$\xibar(z)$, consistent with our measure of $\xibar(\zC)$. Computing the
Thomson optical depth is straightforward once $\xibar(z)$ is in hand -- we
assume HeIII reionization occurs at $z = 3$, and that HeII and hydrogen
reionization occur simultaneously.

\begin{figure*}[htbp]
\begin{center}
\includegraphics[width=0.98\textwidth]{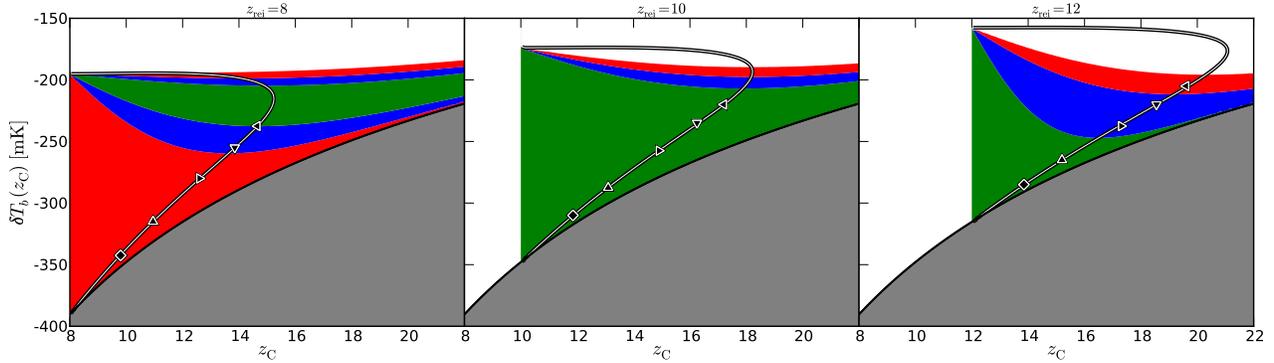}
\caption{Plausibility of an ionization-driven absorption feature assuming
\textit{tanh} models of reionization with $\zrei = 8,10,$ and $12$ from left
to right. The filled contours denote measures of $\dTb(z)$ (and thus $\xibar$
assuming an adiabatically-cooling Universe) which correspond to ionization
histories consistent with WMAP 9 values of $\tau_e$ \citep{Bennett2012} at the
1, 2, and 3-$\sigma$ level (green, blue, and red, respectively). However,
while the filled contours denote plausible reionization scenarios, not all of
them would induce a turning point in the global 21-cm signal. The white
contour denotes $(z, \dTb)$ pairs where Equation \ref{eq:TurningPoint} is
satisfied exactly, meaning $(z, \dTb)$ points lying within the filled contours
\textit{and} along the white contour mark locations where turning point C
would be a produced by ionization and also be consistent with the CMB
constraint. Symbols denote \textit{tanh} models with $\Delta \zrei = 1, 2, 4,
6, 8$ (diamond, upward/rightward/downward/leftward-facing triangles,
respectively). Values of $\Delta \zrei \leq 7.9$ are consistent with the most conservative (model-dependent) constraints from South Pole Telescope \citep[via the kinetic Sunyaev-Zeldovich (SZ) effect;][]{Zahn2012}, which assume no prior knowledge of angular correlations in the cosmic infrared background and thermal SZ power.}
\label{fig:tpC_ion}
\end{center}
\end{figure*}

At a turning point, however, Equation \ref{eq:TurningPoint} must
also be satisfied. This results in a unique track through $(z, \dTb)$ space
corresponding to values of $\zC$ and $\dTb(\zC)$ that are consistent with both
$\xibar(\zC)$ and its time derivative for a given \textit{tanh} model. Figure
\ref{fig:tpC_ion} shows the joint ionization and 21-cm histories consistent
with WMAP 9 constraints on $\tau_e$ \citep{Bennett2012}.

This technique is limited because it assumes a functional form for the
ionization history that may be incorrect, in addition to the fact that we are
only using two points in the fit -- the first being $\zrei$, at which point
$\xibar = 0.5$ (by definition), and the second being $\xibar(\zC)$ as inferred
from $\dTb(\zC)$. However, it does show that reasonable reionization scenarios
could produce turning point C, although at later times (lower redshifts) than
typical models (where turning point C is a byproduct of heating) predict.

%%
% Turning Point D
%%
\subsection{Turning Point D: Reionization} \label{sec:D}
In principle, turning point D could be due to a sudden decline in the $\Lya$
background intensity, which would cause $\TS$ to decouple from $\TK$ and
re-couple to the CMB. Alternatively, turning point D could occur if
heating subsided enough for the Universe to cool back down to the
CMB temperature. However, the more plausible scenario is that coupling
continues between $\TS$ and $\TK$, heating persists, and the signal
``saturates,'' i.e. $1 - \Tcmb/\TS \approx 1$, in which case the brightness
temperature is a direct proxy for the volume filling factor of HII
regions\footnote{If the signal is not yet saturated, a measurement of turning
point D instead yields an upper limit on $\xibar$.}.

If saturated, Equation \ref{eq:TurningPoint} becomes
\begin{equation}
    \frac{\xibar}{1-\xibar} \frac{d\log \xibar}{d\log t} \simeq \left(\frac{\Tcmb}{\TK}\right) \frac{d\log \TK}{d\log t} - \frac{1}{3} . \label{eq:Saturated}
\end{equation}
Even in the saturated regime, the first term on the right-hand side cannot be discarded since we have assumed nothing about $d\log \TS / d\log t$.

Many authors have highlighted the 21-cm emission signal as a probe of the
ionization history during the EoR \citep[e.g.][]{Pritchard2010b,Morandi2012}.
Rather than dwell on it, we simply note that if 21-cm measurements of the
EoR signal are accompanied by independent measures of $\xibar$, in principle
one could glean insights into the thermal history from turning point D as
well.

%%
% DISCUSSION
%%
\section{DISCUSSION} \label{sec:Discussion}

\subsection{A Shift in Methodology}
The redshifted 21-cm signal has been studied by numerous authors in the last
10-15 years. Efforts have concentrated on identifying probable sources of
$\Lya$, Lyman-continuum, and X-ray photons at high-$z$, and then solving for
their combined influence on the thermal and ionization state of gas
surrounding individual objects \citep[e.g.][]{Madau1997, Thomas2008,Chen2008,Venkatesan2011}, or the impact of populations of sources on the 
the global properties of the IGM
\citep[e.g.][]{Choudhury2005,Furlanetto2006,Pritchard2010a}. It has been cited
as a probe of the first stars \citep{Barkana2004}, stellar-mass black holes
and active galactic nuclei
\citep[e.g.][]{Mirabel2011,McQuinn2012b,Tanaka2012,Fragos2013,Mesinger2013}, which primarily influence the thermal history through X-ray heating, but could contribute non-negligibly to reionization \citep[e.g.][]{Dijkstra2004,Pritchard2010b,Morandi2012}. More
recently, more subtle effects have come into focus, such as the relative velocity-difference between baryons
and dark-matter, which delays the formation of the first luminous objects
\citep{Tseliakhovich2010,McQuinn2012,Fialkov2012}.

Forward modeling of this sort, where the input is a set of astrophysical
parameters and the output is a synthetic global 21-cm spectrum, is valuable
because it 1) identifies the processes that most affect the signal, 2) has so
far shown that a 21-cm signal should exist given reasonable models for early
structure formation, and 3) that the signal exhibits the same qualitative
features over a large subset of parameter space. However, this methodology
yields no information about how unique a given model is.

We have taken the opposite approach. Rather than starting from an
astrophysical model and computing the resulting 21-cm spectrum, we begin with
an arbitrary signal characterized by its extrema, and identify the IGM
properties that would be consistent with its observation. The advantage is
that 1) we have a mathematical basis to accompany our intuition about which
physical processes give rise to each feature of the signal, 2) we can see how
reliably IGM properties can be constrained given a perfect measurement of the
signal, and 3) we can predict which models will be degenerate without even
computing a synthetic 21-cm spectrum. 

% EXAMPLE INTERPRETATION
\newpage
\subsection{An Example History} \label{sec:ExampleHistory}
In our analysis, we have found that the 21-cm signal provides more than coarse
estimates of when the first stars and black holes form. Turning points B, C,
and D constrain (quantitatively) the background $\Lya$ intensity, cumulative
energy deposition, and mean ionized fraction, respectively, as well as their
time derivatives, as summarized in Table \ref{tab:SignalFeatures}. For
concreteness, we will now revisit each feature of the signal for an assumed
realization of the 21-cm spectrum, and demonstrate how each can be interpreted
in terms of model-independent IGM properties.

%%
% TABLE: Features of Global Signal
%%
\begin{deluxetable*}{clclccc}
\tabletypesize{\scriptsize}
\tablecaption{Features of the global 21-cm signal}
\tablecolumns{7}
\tablehead{Feature & Measurement & Assumptions & Yield & \S & Equations & Figures} 
\startdata
B & $\zB$ & ... & lower limit on redshift of first star formation & \ref{sec:B} & ... & ... \\
B & $\dTb(\zB)$ &  $\xibar = \eheat = 0$ & $\Jhat(\zB)$, $\Jhat^{\prime}(\zB)$ & \ref{sec:B} & \ref{eq:dTb}-\ref{eq:Jalpha}, \ref{eq:TurningPoint}-\ref{eq:dlogxc} &  \ref{fig:tpB} \\
\hline
C & $\zC$ & ... & upper limit on $\eheat(\zC)$ & \ref{sec:C} & \ref{eq:TemperatureSolution} & \ref{fig:tpC_heat_constraints} \\
C & $\zC$ & $\xibar = 0$ & lower limit on redshift of first X-ray source formation & \ref{sec:C} & ... & ... \\ 
C & $\zC$ & $\xibar = 0$ & lower limit on $\eheat(\zC)$ & \ref{sec:C} & \ref{eq:ThermalEvolution} & \ref{fig:thermal_evolution}, \ref{fig:tpC_heat_constraints} \\  
C & $\dTb(\zC)$ & ... & improved upper limit on $\eheat(\zC)$ & \ref{sec:C} & \ref{eq:dTb}, \ref{eq:TurningPoint}, \ref{eq:ThermalEvolution}, \ref{eq:TemperatureSolution} & \ref{fig:tpC_heat_constraints} \\
C  & $\dTb(\zC)$ & $\eheat = 0$ & rule out reionization scenario? & \ref{sec:Cion}  & \ref{eq:TurningPoint} & \ref{fig:tpC_ion} \\
\hline
transition & $\ztrans$ & $\TS = \TK$ & upper limit on $\int \eheat dt$ & \ref{sec:trans} & \ref{eq:TemperatureSolution} & \ref{fig:slope_constraints} \\ 
transition & $\frac{d}{d\nu}\left[\delta T_b\right](\ztrans)$ & $\TS = \TK$ & lower limit on $\eheat(\ztrans)$ & \ref{sec:trans} & \ref{eq:TransitionSlope} & \ref{fig:slope_constraints}  \\
\hline 
D & $\zD$ & ... & start of EoR & \ref{sec:D} & ... & ... \\
D & $\dTb(\zD)$ & ... & upper limit on $\xibar(\zD)$  & \ref{sec:D} & \ref{eq:dTb} & ... \\
D & $\dTb(\zD)$ & $\TS = \TK \gg \Tcmb$ & $\xibar(\zD)$, joint constraint on $\xibar^{\prime}(\zD)$, $\TK(\zD)$, and $\TK^{\prime}(\zD)$ & \ref{sec:D} & \ref{eq:dTb}, \ref{eq:TurningPoint}, \ref{eq:Saturated} & ...
\enddata
\tablecomments{Constraints on IGM properties from critical points in the global 21-cm signal. Each block focuses on a single feature of the signal (denoted in column \#1) and from left to right reports how a given measurement (column \#2; e.g. the feature's redshift, $z$) under some set of assumptions (column \#3) would be interpreted (column \#4). The corresponding section of the text, as well as any equations and figures relevant to the given feature are listed in columns 5, 6, and 7, respectively. Within each block, elements appear in order of increasing complexity (in terms of the measurement difficulty and number of assumptions) from top to bottom.}
\label{tab:SignalFeatures}
\end{deluxetable*}

We will assume the same realization of the signal as is shown in Figure
\ref{fig:global_signal}, with turning points B, C, and D at $(z, \dTb /
\mathrm{mK})$ of \tpB, \tpC, and \tpD, respectively, and absorption-emission
transition at $\ztrans=15$, $d(\dTb)/d\nu = 4.3 \ \mathrm{mK} \
\mathrm{MHz}^{-1}$. At a glance, the 21-cm realization shown in Figure
\ref{fig:global_signal} indicates that the Universe's first stars form at $z
\gtrsim 30$, the first black holes form at $z \gtrsim 21$, and that
reionization has begun by $z \gtrsim 13.5$. Global feedback models such as
those presented in \citet{Tanaka2012} are inconsistent with this realization
of the signal, as they predict $\TK > \Tcmb$ at $z \gtrsim 20$.

More quantitatively, from Figure \ref{fig:tpB} we have an upper limit on the
$\Lya$ background intensity of $\Jhat(\zB) \geq 10^{-11.1}
\intensityunitsnumber$ and its time rate-of-change, $d\log \Jhat / d\log t
\simeq 11.2$. Moving on to turning point C (Figure
\ref{fig:tpC_heat_constraints}), the kinetic temperature is constrained
between $9 \lesssim \TK / \mathrm{K} \lesssim 16$, meaning the cumulative
energy deposition must be $\int \eheat dt \leq 10^{51.9} \mathrm{erg} \
\mathrm{cMpc}^{-3}$. In the absence of any ionization, a minimum heating rate
density of $\eheat \geq 10^{36.1} \coheatingdensity$ is required to produce
turning point C, and a maximum of $\eheat \leq 10^{38.2} \coheatingdensity$ is
imposed given the existence of the absorption feature.

The slope of the signal as it crosses $\dTb = 0$ is $\dTb^{\prime} = 4.3
\mathrm{mK} \ \mathrm{MHz}^{-1}$, corresponding to a lower limit on the
heating rate density of $\eheat \geq 10^{37.6} \coheatingdensity$ (Figure
\ref{fig:slope_constraints}). Finally, at turning point D, the ionized
fraction must be $\xibar \leq 0.24$ (Eq. \ref{eq:dTb} when $\TS
>> \Tcmb$). An ionization-driven turning point C can be ruled out by Figure
\ref{fig:tpC_ion}, since the amount of ionization required to produce $(\zC,
\dTb(\zC)) =$ \tpC leads to $\tau_e$ values inconsistent with WMAP at the
$> 3\sigma$ level, for \textit{tanh} models with $8 \leq \zrei \leq 12$.

With limits on $\Jhat$, $\eheat$, $\xibar$, and their derivatives, the next
step is to determine how each quantity relates to astrophysical quantities.
Typically, models for the global 21-cm signal relate the emissivity of the
Universe to the cosmic star-formation rate density (SFRD) via simple scalings
of the form $\hat{\upepsilon}_{i,\nu}(z) \propto f_i \rhostardot(z)
I_{\nu}$ \citep[e.g.][]{Furlanetto2006,Pritchard2010a}, in
which case the parameters of interest are $f_i$, which converts a star
formation rate into a bolometric energy output in band $i$ (generally split
between $\Lya$, soft-UV, and X-ray photons), the SFRD itself, $\rhostardot$,
and the spectral energy distribution (SED) of luminous sources being modeled,
$I_{\nu}$.

Given that soft-UV photons have very short mean-free-paths in a neutral
medium, a determination of $d\log \xibar / d\log t$ is likely to be an
accurate tracer of the soft-UV ionizing emissivity of the Universe, $\eion$.
However, the same is not true of photons emitted between $\Lyn$ resonances and hard X-ray photons, which can travel
large distances before being absorbed, where they predominantly contribute to
Wouthuysen-Field coupling and heating, respectively. Because of this,
translating $\Jhat$ and $\eheat$ measurements to their corresponding
emissivities, $\ealpha$ and $\eX$, is non-trivial. In general,
the accuracy with which one can convert $\Jhat$ ($\eheat$) to $\ealpha$
($\eX$) depends on the redshift-evolution of the co-moving bolometric
luminosity and the SED of sources, $I_{\nu}$. 

For a zeroth order estimate, we will assume that sources have a flat spectrum between the $\Lya$ resonance and the Lyman limit, and neglect ``injected photons,'' i.e. those that redshift into higher a $\Lyn$ resonance and (possibly) cascade through the $\Lya$ resonance. If $\ealpha \propto N_{\alpha} \rhostardot$, where $N_{\alpha}$ is the number of photons emitted between $\nuLya \leq \nu \leq \nuLL$ per baryon, then
\begin{equation}
    \rhostardot(z) \approx 10^{-5} \left(\frac{9690}{N_{\alpha}}\right) \left(\frac{J_{\alpha}}{J_{21}} \right) \left(\frac{1 + z}{30} \right)^{-1/2} \Msun \ \mathrm{yr}^{-1} \ \mathrm{cMpc}^{-3}  
\end{equation}
where we've scaled $N_{\alpha}$ to a value appropriate for low-mass PopII stars \citep{Barkana2004}.

Similarly, if we assume that a fraction $\fXh = 0.2$ of the X-ray emissivity is deposited as heat \citep[appropriate for the $E \gtrsim 0.1$ keV limit in a neutral medium;][]{Shull1985}, and normalize by the local $L_X$-SFR relationship \citep[e.g.][who found $L_{0.5-8 \mathrm{keV}} = 2.61 \times 10^{39} \ \mathrm{erg} \ \mathrm{s}^{-1} \ (\Msun \ \mathrm{yr}^{-1})$]{Mineo2012}, we have
\begin{align}
    \rhostardot(z) & \approx 2 \times 10^{-2} \fX^{-1} \left(\frac{0.2}{f_{\mathrm{X,h}}} \right) \nonumber \\
    & \times \left(\frac{\eheat}{10^{37} \ \coheatingdensity} \right) \ \Msun \ \mathrm{yr}^{-1} \ \mathrm{cMpc}^{-3}
\end{align}    
where we subsume all uncertainty in the normalization between $L_X$ and $\rhostardot$, the SED of X-ray sources, and radiative transfer effects into the factor $\fX$.

If these approximate treatments are sufficient, then measures of $J_{\alpha}$
provide 2D constraints on $\rhostardot$ and $N_{\alpha}$, and measures of
$\eheat$ constrain $\rhostardot$ and $\fX$\footnote{Here we have assumed
that high-mass X-ray binaries are the only source of X-rays, when in reality
the heating may be induced by a variety of sources. Other candidates include
X-rays from ``miniquasars'' \citep[e.g.][]{Kuhlen2005}, inverse Compton
scattered CMB photons off high energy electrons accelerated in supernovae
remnants \citep{Oh2001}, or shock heating \citep[e.g.][]{Gnedin2004,Furlanetto2004}.}.
However, given the long mean free paths of X-rays and photons in the
$\nuLya \leq \nu \leq \nuLL$ band, the estimates above are likely to be
inadequate. It is the primary goal of a forthcoming paper (Mirocha et al., in
prep) to characterize uncertainties in these estimates that arise due to two
major unknowns: 1) redshift evolution in the ionizing emissivity of UV and
X-ray sources, and 2) their spectral energy distributions.

% SYNERGIES
\subsection{Synergies with Upcoming Facilities}
The prospects for synergies are most promising for turning point D, which is
predicted to occur at $z \lesssim 15$, coinciding with the JWST window and
current and upcoming campaigns to measure the 21-cm power spectrum. JWST will
probe the high-$z$ galaxy population even more sensitively than HST
\citep[e.g.][]{Robertson2013}, which may allow degeneracies between the
star-formation history and other parameters to be broken (e.g. the $f_i$
normalization factors). However, our focus in this paper is on
model-independent quantities -- the issue of degeneracy among astrophysical
parameters will be discussed in Paper II.

In terms of model-independent quantities, current and upcoming facilities will
benefit global 21-cm measurements by constraining the ionization history. For
example, one can constrain $\xibar(z)$ via observations of $\Lya$-emitters
\citep[LAEs, e.g.][]{Malhotra2006,McQuinn2007,Mesinger2008}, the CMB through
$\tau_e$ and the kinetic Sunyaev-Zeldovich effect \citep{Zahn2012}, or via
measurements of the 21-cm power spectrum, which reliably peaks when $\xibar
\simeq 0.5 $\citep{Lidz2008}. However, like the global signal, power spectrum
measurements yield upper limits on $\xibar$, since they assume $\TS \gg
\Tcmb$, which may not be the case. Constraints from LAEs require no such
assumption, and instead set lower limits on $\xibar$, since our ability to see
$\Lya$ emission from galaxies at high-$z$ depends on the \textit{minimum} size
of an HII region required for $\Lya$ photons to escape. Limits on $\xibar(z)$
out to $z \sim 10-15$ would yield a prediction for the amplitude of turning
point D, which, in conjunction with a global 21-cm measurement could validate
or invalidate the $\TS \gg \Tcmb$ assumption often adopted for EoR work. In
addition, one could determine if ionization-driven absorption features are
even remotely feasible (\S\ref{sec:tpC_ion}).

% CAVEATS
\subsection{Caveats}
Simple models for the global 21-cm signal rely on the assumption that the IGM
is well approximated as a two-phase medium, one phase representing HII
regions, and the other representing the bulk IGM. As reionization progresses,
the distinction between these two phases will become tenuous, owing to a
warming and increasingly ionized IGM whose properties differ little from an
HII region. Even prior to reionization the global approximation may be
inadequate depending on the distribution of luminous sources. If exceedingly
rare sources dominate ionization and heating, we would require a more detailed
treatment \citep[a problem recently addressed in the context of helium
reionization by][]{Davies2012}.

Eventually, simple models must also be calibrated by more sophisticated
simulations. This has been done to some extent already in the context of 21-cm
fluctuations, with good agreement so far between semi-analytic and numerical
models \citep{Zahn2011}. However, analogous comparisons for the global signal
have yet to be performed rigorously. The limiting factor is that a large
volume must be simulated in order to avoid cosmic variance, but the spatial
resolution required to simultaneously resolve the first galaxies becomes
computationally restrictive.

Finally, though we included an analysis of the absorption-emission transition
point, $\ztrans$, in truth, the slope measured from this feature will be
correlated with the positions of the turning points. The most
promising foreground removal studies rely on parameterizing the signal as a
simple function (e.g. spline), meaning the slope at $\ztrans$ is completely
determined by the positions of the turning points and the function used to
represent the astrophysical signal.

%%
% CONCLUSIONS
%%
\section{CONCLUSIONS}
In this paper we have addressed one tier of the 21-cm interpretation problem:
identifying the physical properties of the IGM that can be constrained
uniquely from a measurement of the all-sky 21-cm signal. Our main conclusions
are:
\begin{itemize}
    \item The first feature of the global signal, turning point B, provides a lower limit on the redshift at which the Universe's first stars formed. But, more quantitatively, its position in $(z, \dTb)$ space measures the background $\Lya$ intensity, $\Jhat$, and its time derivative, respectively, assuming a neutral, adiabatically-cooling medium.
    \item The absorption feature, turning point C, is most likely a probe of accretion onto compact objects considering the $\tau_e$ constraint from the CMB. As a result, it provides a lower limit on the redshift when the first X-ray emitting objects formed. Even if the magnitude of the absorption trough cannot be accurately measured, a determination of $\zC$ alone sets strong upper and lower limits on the heating rate density of the Universe, $\eheat(\zC)$. If the absorption feature is deep ($\dTb(\zC) \lesssim -200$ mK) and occurs late ($z \lesssim 15$), it could be a byproduct of reionization.
    \item The final feature, turning point D, indicates the start of the EoR, and traces the mean ionized fraction of the Universe and its time derivative. In general, it also depends on the spin-temperature evolution, though it is expected that at this stage the signal is fully saturated. Without independent constraints on the thermal history, $\dTb(\zD)$ provides an upper limit on the mean ionized fraction, $\overline{x}_i$.
\end{itemize}

In general, the relationship between IGM diagnostics (such as $\Jhat$ and $\eheat$) and the properties of the astrophysical sources themselves (like $\rhostardot$, $\Nalpha$, and $\fX$) is expected to be complex. In a forthcoming paper, we compare simple analytic arguments (e.g. those used in \S\ref{sec:ExampleHistory}) with the results of detailed numerical solutions to the cosmological radiative transfer equation in order to assess how accurately the global 21-cm signal can constrain the Universe's luminous sources.

The authors thank the anonymous referee, whose suggestions helped
improve the quality of this manuscript, and acknowledge the LUNAR
consortium\footnote{http://lunar.colorado.edu}, headquartered at the
University of Colorado, which is funded by the NASA Lunar Science Institute
(via Cooperative Agreement NNA09DB30A) to investigate concepts for
astrophysical observatories on the Moon.

\bibliography{references}

\begin{thebibliography}{104}
\expandafter\ifx\csname natexlab\endcsname\relax\def\natexlab#1{#1}\fi

\bibitem[{{Abel} {et~al.}(2002){Abel}, {Bryan}, \& {Norman}}]{Abel2002b}
{Abel}, T., {Bryan}, G.~L., \& {Norman}, M.~L. 2002, Science, 295, 93

\bibitem[{{Ali-Ha{\"i}moud} \& {Hirata}(2010)}]{AliHamoud2010}
{Ali-Ha{\"i}moud}, Y. \& {Hirata}, C.~M. 2010, \prd, 82, 063521

\bibitem[{{Allison} \& {Dalgarno}(1969)}]{Allison1969}
{Allison}, A.~C. \& {Dalgarno}, A. 1969, \apj, 158, 423

\bibitem[{Barkana \& Loeb(2005)}]{Barkana2004}
Barkana, R. \& Loeb, A. 2005, \apj, 626, 1

\bibitem[{{Begelman} {et~al.}(2008){Begelman}, {Rossi}, \&
  {Armitage}}]{Begelman2008}
{Begelman}, M.~C., {Rossi}, E.~M., \& {Armitage}, P.~J. 2008, \mnras, 387, 1649

\bibitem[{{Begelman} {et~al.}(2006){Begelman}, {Volonteri}, \&
  {Rees}}]{Begelman2006}
{Begelman}, M.~C., {Volonteri}, M., \& {Rees}, M.~J. 2006, \mnras, 370, 289

\bibitem[{Bennett {et~al.}(2012)Bennett, Larson, Weiland, Jarosik, Hinshaw,
  Odegard, Smith, Hill, Gold, Halpern, Komatsu, Nolta, Page, Spergel, Wollack,
  Dunkley, Kogut, Limon, Meyer, Tucker, \& Wright}]{Bennett2012}
Bennett, C.~L. {et~al.} 2012, preprint (astroph/12125225)

\bibitem[{{Bouwens} {et~al.}(2011){Bouwens}, {Illingworth}, {Oesch},
  {Labb{\'e}}, {Trenti}, {van Dokkum}, {Franx}, {Stiavelli}, {Carollo},
  {Magee}, \& {Gonzalez}}]{Bouwens2011}
{Bouwens}, R.~J. {et~al.} 2011, \apj, 737, 90

\bibitem[{{Bowman} {et~al.}(2013){Bowman}, {Cairns}, {Kaplan}, {Murphy},
  {Oberoi}, {Staveley-Smith}, {Arcus}, {Barnes}, {Bernardi}, {Briggs}, {Brown},
  {Bunton}, {Burgasser}, {Cappallo}, {Chatterjee}, {Corey}, {Coster},
  {Deshpande}, {deSouza}, {Emrich}, {Erickson}, {Goeke}, {Gaensler},
  {Greenhill}, {Harvey-Smith}, {Hazelton}, {Herne}, {Hewitt},
  {Johnston-Hollitt}, {Kasper}, {Kincaid}, {Koenig}, {Kratzenberg}, {Lonsdale},
  {Lynch}, {Matthews}, {McWhirter}, {Mitchell}, {Morales}, {Morgan}, {Ord},
  {Pathikulangara}, {Prabu}, {Remillard}, {Robishaw}, {Rogers}, {Roshi},
  {Salah}, {Sault}, {Shankar}, {Srivani}, {Stevens}, {Subrahmanyan}, {Tingay},
  {Wayth}, {Waterson}, {Webster}, {Whitney}, {Williams}, {Williams}, \&
  {Wyithe}}]{Bowman2013}
{Bowman}, J.~D. {et~al.} 2013, Publications of the Astronomical Society of
  Australia, 30, 31

\bibitem[{{Bowman} \& {Rogers}(2010)}]{Bowman2010}
{Bowman}, J.~D. \& {Rogers}, A.~E.~E. 2010, \nat, 468, 796

\bibitem[{{Bromm} {et~al.}(1999){Bromm}, {Coppi}, \& {Larson}}]{Bromm1999}
{Bromm}, V., {Coppi}, P.~S., \& {Larson}, R.~B. 1999, \apjl, 527, L5

\bibitem[{Bromm {et~al.}(2001)Bromm, Kudritzki, \& Loeb}]{Bromm2001}
Bromm, V., Kudritzki, R.~P., \& Loeb, A. 2001, \apj, 552, 464

\bibitem[{{Bromm} {et~al.}(2009){Bromm}, {Yoshida}, {Hernquist}, \&
  {McKee}}]{Bromm2009}
{Bromm}, V., {Yoshida}, N., {Hernquist}, L., \& {McKee}, C.~F. 2009, \nat, 459,
  49

\bibitem[{{Burns} {et~al.}(2012){Burns}, {Lazio}, {Bale}, {Bowman}, {Bradley},
  {Carilli}, {Furlanetto}, {Harker}, {Loeb}, \& {Pritchard}}]{Burns2012}
{Burns}, J.~O. {et~al.} 2012, Advances in Space Research, 49, 433

\bibitem[{{Carilli} {et~al.}(2004){Carilli}, {Furlanetto}, {Briggs}, {Jarvis},
  {Rawlings}, \& {Falcke}}]{Carilli2004}
{Carilli}, C.~L., {Furlanetto}, S., {Briggs}, F., {Jarvis}, M., {Rawlings}, S.,
  \& {Falcke}, H. 2004, New Astronomy Reviews, 48, 1029

\bibitem[{Chen {et~al.}(2008)Chen, Chen, Miralda-Escud{\'e}, \&
  Miralda-Escud{\'e}}]{Chen2008}
Chen, X., Chen, X., Miralda-Escud{\'e}, J., \& Miralda-Escud{\'e}, J. 2008,
  \apj, 684, 18

\bibitem[{{Chen} \& {Miralda-Escud{\'e}}(2004)}]{Chen2004}
{Chen}, X. \& {Miralda-Escud{\'e}}, J. 2004, \apj, 602, 1

\bibitem[{{Chluba} \& {Thomas}(2011)}]{Chluba2011}
{Chluba}, J. \& {Thomas}, R.~M. 2011, \mnras, 412, 748

\bibitem[{{Choudhury} \& {Ferrara}(2005)}]{Choudhury2005}
{Choudhury}, T.~R. \& {Ferrara}, A. 2005, \mnras, 361, 577

\bibitem[{Ciardi {et~al.}(2010)Ciardi, Salvaterra, \& Di~Matteo}]{Ciardi2010}
Ciardi, B., Salvaterra, R., \& Di~Matteo, T. 2010, \mnras, 401, 2635

\bibitem[{{Coe} {et~al.}(2013){Coe}, {Zitrin}, {Carrasco}, {Shu}, {Zheng},
  {Postman}, {Bradley}, {Koekemoer}, {Bouwens}, {Broadhurst}, {Monna}, {Host},
  {Moustakas}, {Ford}, {Moustakas}, {van der Wel}, {Donahue}, {Rodney},
  {Ben{\'{\i}}tez}, {Jouvel}, {Seitz}, {Kelson}, \& {Rosati}}]{Coe2013}
{Coe}, D. {et~al.} 2013, \apj, 762, 32

\bibitem[{{Davies} \& {Furlanetto}(2012)}]{Davies2012}
{Davies}, F.~B. \& {Furlanetto}, S.~R. 2012, preprint (astroph/12094900)

\bibitem[{Dijkstra {et~al.}(2004)Dijkstra, Haiman, \& Loeb}]{Dijkstra2004}
Dijkstra, M., Haiman, Z., \& Loeb, A. 2004, \apj, 613, 646

\bibitem[{{Dunkley} {et~al.}(2009){Dunkley}, {Komatsu}, {Nolta}, {Spergel},
  {Larson}, {Hinshaw}, {Page}, {Bennett}, {Gold}, {Jarosik}, {Weiland},
  {Halpern}, {Hill}, {Kogut}, {Limon}, {Meyer}, {Tucker}, {Wollack}, \&
  {Wright}}]{Dunkley2009}
{Dunkley}, J. {et~al.} 2009, \apjs, 180, 306

\bibitem[{{Ellis} {et~al.}(2013){Ellis}, {McLure}, {Dunlop}, {Robertson},
  {Ono}, {Schenker}, {Koekemoer}, {Bowler}, {Ouchi}, {Rogers}, {Curtis-Lake},
  {Schneider}, {Charlot}, {Stark}, {Furlanetto}, \& {Cirasuolo}}]{Ellis2013}
{Ellis}, R.~S. {et~al.} 2013, \apjl, 763, L7

\bibitem[{Fan(2006)}]{Fan2006}
Fan, X. 2006, New Astronomy Reviews, 50, 665

\bibitem[{Fialkov {et~al.}(2012)Fialkov, Barkana, Tseliakhovich, \&
  Hirata}]{Fialkov2012}
Fialkov, A., Barkana, R., Tseliakhovich, D., \& Hirata, C.~M. 2012, \mnras,
  424, 1335

\bibitem[{{Field}(1958)}]{Field1958}
{Field}, G.~B. 1958, Proceedings of the IRE, 46, 240

\bibitem[{Fragos {et~al.}(2013)Fragos, Lehmer, Naoz, Zezas, \&
  Basu-Zych}]{Fragos2013}
Fragos, T., Lehmer, B.~D., Naoz, S., Zezas, A., \& Basu-Zych, A.~R. 2013,
  preprint (astroph/13061405)

\bibitem[{Furlanetto(2006)}]{Furlanetto2006}
Furlanetto, S.~R. 2006, \mnras, 371, 867

\bibitem[{{Furlanetto} \& {Furlanetto}(2007)}]{FurlanettoFurlanetto2007a}
{Furlanetto}, S.~R. \& {Furlanetto}, M.~R. 2007, \mnras, 374, 547

\bibitem[{Furlanetto \& Furlanetto(2007)}]{FurlanettoFurlanetto2007b}
Furlanetto, S.~R. \& Furlanetto, M.~R. 2007, \mnras, 379, 130

\bibitem[{Furlanetto \& Loeb(2004)}]{Furlanetto2004}
Furlanetto, S.~R. \& Loeb, A. 2004, \apj, 611, 642

\bibitem[{Furlanetto {et~al.}(2006{\natexlab{a}})Furlanetto, Oh, \&
  Briggs}]{FurlanettoOhBriggs2006}
Furlanetto, S.~R., Oh, S.~P., \& Briggs, F.~H. 2006{\natexlab{a}}, Physics
  Reports, 433, 181

\bibitem[{Furlanetto {et~al.}(2006{\natexlab{b}})Furlanetto, Oh, \&
  Pierpaoli}]{FurlanettoOh2006}
Furlanetto, S.~R., Oh, S.~P., \& Pierpaoli, E. 2006{\natexlab{b}}, Physical
  Review D, 74, 103502

\bibitem[{{Furlanetto} \& {Pritchard}(2006)}]{FurlanettoPritchard2006}
{Furlanetto}, S.~R. \& {Pritchard}, J.~R. 2006, \mnras, 372, 1093

\bibitem[{{Gnedin} \& {Shaver}(2004)}]{Gnedin2004}
{Gnedin}, N.~Y. \& {Shaver}, P.~A. 2004, \apj, 608, 611

\bibitem[{{Gunn} \& {Peterson}(1965)}]{Gunn1965}
{Gunn}, J.~E. \& {Peterson}, B.~A. 1965, \apj, 142, 1633

\bibitem[{{Haiman} {et~al.}(2000){Haiman}, {Abel}, \& {Rees}}]{Haiman2000}
{Haiman}, Z., {Abel}, T., \& {Rees}, M.~J. 2000, \apj, 534, 11

\bibitem[{{Haiman} {et~al.}(1996){Haiman}, {Thoul}, \& {Loeb}}]{Haiman1996}
{Haiman}, Z., {Thoul}, A.~A., \& {Loeb}, A. 1996, \apj, 464, 523

\bibitem[{{Harker} {et~al.}(2010){Harker}, {Zaroubi}, {Bernardi}, {Brentjens},
  {de Bruyn}, {Ciardi}, {Jeli{\'c}}, {Koopmans}, {Labropoulos}, {Mellema},
  {Offringa}, {Pandey}, {Pawlik}, {Schaye}, {Thomas}, \&
  {Yatawatta}}]{Harker2010}
{Harker}, G. {et~al.} 2010, \mnras, 405, 2492

\bibitem[{{Harker} {et~al.}(2012){Harker}, {Pritchard}, {Burns}, \&
  {Bowman}}]{Harker2012}
{Harker}, G.~J.~A., {Pritchard}, J.~R., {Burns}, J.~O., \& {Bowman}, J.~D.
  2012, \mnras, 419, 1070

\bibitem[{Hirata(2006)}]{Hirata2006}
Hirata, C.~M. 2006, \mnras, 367, 259

\bibitem[{{Hirata} \& {Sigurdson}(2007)}]{Hirata2007}
{Hirata}, C.~M. \& {Sigurdson}, K. 2007, \mnras, 375, 1241

\bibitem[{{Jester} \& {Falcke}(2009)}]{Jester2009}
{Jester}, S. \& {Falcke}, H. 2009, New Astronomy Reviews, 53, 1

\bibitem[{{Johnson}(2010)}]{Johnson2010}
{Johnson}, J.~L. 2010, \mnras, 404, 1425

\bibitem[{Johnson {et~al.}(2009)Johnson, Greif, Bromm, Klessen, \&
  Ippolito}]{Johnson2009}
Johnson, J.~L., Greif, T.~H., Bromm, V., Klessen, R.~S., \& Ippolito, J. 2009,
  \mnras, 399, 37

\bibitem[{{Johnson} {et~al.}(2012){Johnson}, {Whalen}, {Fryer}, \&
  {Li}}]{Johnson2012}
{Johnson}, J.~L., {Whalen}, D.~J., {Fryer}, C.~L., \& {Li}, H. 2012, \apj, 750,
  66

\bibitem[{{Komatsu} {et~al.}(2011){Komatsu}, {Smith}, {Dunkley}, {Bennett},
  {Gold}, {Hinshaw}, {Jarosik}, {Larson}, {Nolta}, {Page}, {Spergel},
  {Halpern}, {Hill}, {Kogut}, {Limon}, {Meyer}, {Odegard}, {Tucker}, {Weiland},
  {Wollack}, \& {Wright}}]{Komatsu2011}
{Komatsu}, E. {et~al.} 2011, \apjs, 192, 18

\bibitem[{{Kuhlen} \& {Madau}(2005)}]{Kuhlen2005}
{Kuhlen}, M. \& {Madau}, P. 2005, \mnras, 363, 1069

\bibitem[{{Larson} {et~al.}(2011){Larson}, {Dunkley}, {Hinshaw}, {Komatsu},
  {Nolta}, {Bennett}, {Gold}, {Halpern}, {Hill}, {Jarosik}, {Kogut}, {Limon},
  {Meyer}, {Odegard}, {Page}, {Smith}, {Spergel}, {Tucker}, {Weiland},
  {Wollack}, \& {Wright}}]{Larson2011}
{Larson}, D. {et~al.} 2011, \apjs, 192, 16

\bibitem[{{Lidz} {et~al.}(2008){Lidz}, {Zahn}, {McQuinn}, {Zaldarriaga}, \&
  {Hernquist}}]{Lidz2008}
{Lidz}, A., {Zahn}, O., {McQuinn}, M., {Zaldarriaga}, M., \& {Hernquist}, L.
  2008, \apj, 680, 962

\bibitem[{Liu {et~al.}(2013)Liu, Pritchard, Tegmark, \& Loeb}]{Liu2013}
Liu, A., Pritchard, J.~R., Tegmark, M., \& Loeb, A. 2013, Physical Review D,
  87, 043002

\bibitem[{Madau {et~al.}(1997)Madau, Meiksin, \& Rees}]{Madau1997}
Madau, P., Meiksin, A., \& Rees, M.~J. 1997, \apj, 475, 429

\bibitem[{Malhotra \& Rhoads(2006)}]{Malhotra2006}
Malhotra, S. \& Rhoads, J.~E. 2006, \apj, 647, L95

\bibitem[{Mcquinn(2012)}]{McQuinn2012b}
Mcquinn, M. 2012, \mnras, 426, 1349

\bibitem[{McQuinn {et~al.}(2007)McQuinn, Hernquist, Zaldarriaga, \&
  Dutta}]{McQuinn2007}
McQuinn, M., Hernquist, L., Zaldarriaga, M., \& Dutta, S. 2007, \mnras, 381, 75

\bibitem[{{McQuinn} \& {O'Leary}(2012)}]{McQuinn2012}
{McQuinn}, M. \& {O'Leary}, R.~M. 2012, \apj, 760, 3

\bibitem[{{Mellema} {et~al.}(2013){Mellema}, {Koopmans}, {Abdalla}, {Bernardi},
  {Ciardi}, {Daiboo}, {de Bruyn}, {Datta}, {Falcke}, {Ferrara}, {Iliev},
  {Iocco}, {Jeli{\'c}}, {Jensen}, {Joseph}, {Labroupoulos}, {Meiksin},
  {Mesinger}, {Offringa}, {Pandey}, {Pritchard}, {Santos}, {Schwarz},
  {Semelin}, {Vedantham}, {Yatawatta}, \& {Zaroubi}}]{Mellema2013}
{Mellema}, G. {et~al.} 2013, Experimental Astronomy

\bibitem[{Mesinger {et~al.}(2013)Mesinger, Ferrara, \& Spiegel}]{Mesinger2013}
Mesinger, A., Ferrara, A., \& Spiegel, D.~S. 2013, \mnras

\bibitem[{Mesinger \& Furlanetto(2008)}]{Mesinger2008}
Mesinger, A. \& Furlanetto, S.~R. 2008, \mnras, 386, 1990

\bibitem[{{Mineo} {et~al.}(2012){Mineo}, {Gilfanov}, \& {Sunyaev}}]{Mineo2012}
{Mineo}, S., {Gilfanov}, M., \& {Sunyaev}, R. 2012, \mnras, 419, 2095

\bibitem[{Mirabel {et~al.}(2011)Mirabel, Dijkstra, Laurent, Loeb, \&
  Pritchard}]{Mirabel2011}
Mirabel, I.~F., Dijkstra, M., Laurent, P., Loeb, A., \& Pritchard, J.~R. 2011,
  preprint (astroph/11021891), astro-ph.CO

\bibitem[{{Morandi} \& {Barkana}(2012)}]{Morandi2012}
{Morandi}, A. \& {Barkana}, R. 2012, \mnras, 424, 2551

\bibitem[{{Mortlock} {et~al.}(2011){Mortlock}, {Warren}, {Venemans}, {Patel},
  {Hewett}, {McMahon}, {Simpson}, {Theuns}, {Gonz{\'a}les-Solares}, {Adamson},
  {Dye}, {Hambly}, {Hirst}, {Irwin}, {Kuiper}, {Lawrence}, \&
  {R{\"o}ttgering}}]{Mortlock2011}
{Mortlock}, D.~J. {et~al.} 2011, \nat, 474, 616

\bibitem[{Oesch {et~al.}(2010)Oesch, Bouwens, Illingworth, Carollo, Franx,
  Labb{\'e}, Magee, Stiavelli, Trenti, \& Van~Dokkum}]{Oesch2010}
Oesch, P.~A. {et~al.} 2010, \apjl, 709, L16

\bibitem[{{Oesch} {et~al.}(2012){Oesch}, {Bouwens}, {Illingworth}, {Gonzalez},
  {Trenti}, {van Dokkum}, {Franx}, {Labb{\'e}}, {Carollo}, \&
  {Magee}}]{Oesch2012}
{Oesch}, P.~A. {et~al.} 2012, \apj, 759, 135

\bibitem[{Oh(2001)}]{Oh2001}
Oh, S.~P. 2001, \apj, 553, 499

\bibitem[{{Paciga} {et~al.}(2013){Paciga}, {Albert}, {Bandura}, {Chang},
  {Gupta}, {Hirata}, {Odegova}, {Pen}, {Peterson}, {Roy}, {Shaw}, {Sigurdson},
  \& {Voytek}}]{Paciga2013}
{Paciga}, G. {et~al.} 2013, \mnras, 433, 639

\bibitem[{{Parsons} {et~al.}(2010){Parsons}, {Backer}, {Foster}, {Wright},
  {Bradley}, {Gugliucci}, {Parashare}, {Benoit}, {Aguirre}, {Jacobs},
  {Carilli}, {Herne}, {Lynch}, {Manley}, \& {Werthimer}}]{Parsons2010}
{Parsons}, A.~R. {et~al.} 2010, \aj, 139, 1468

\bibitem[{Parsons {et~al.}(2013)Parsons, Liu, Aguirre, Ali, Bradley, Carilli,
  DeBoer, Dexter, Gugliucci, Jacobs, Klima, MacMahon, Manley, Moore, Pober,
  Stefan, \& Walbrugh}]{Parsons2013}
Parsons, A.~R. {et~al.} 2013, preprint (astroph/1304.4991)

\bibitem[{{Peebles}(1993)}]{Peebles1993}
{Peebles}, P.~J.~E. 1993, {Principles of Physical Cosmology} (Princeton
  University Press)

\bibitem[{{Petrovic} \& {Oh}(2011)}]{Petrovic2011}
{Petrovic}, N. \& {Oh}, S.~P. 2011, \mnras, 413, 2103

\bibitem[{Pritchard \& Furlanetto(2007)}]{Pritchard2007}
Pritchard, J.~R. \& Furlanetto, S.~R. 2007, \mnras, 376, 1680

\bibitem[{Pritchard \& Loeb(2010)}]{Pritchard2010a}
Pritchard, J.~R. \& Loeb, A. 2010, Physical Review D, 82, 23006

\bibitem[{Pritchard {et~al.}(2010)Pritchard, Loeb, \& Wyithe}]{Pritchard2010b}
Pritchard, J.~R., Loeb, A., \& Wyithe, J. S.~B. 2010, \mnras, 408, 57

\bibitem[{Ricotti {et~al.}(2005)Ricotti, Ostriker, \& Gnedin}]{Ricotti2005}
Ricotti, M., Ostriker, J.~P., \& Gnedin, N.~Y. 2005, \mnras, 357, 207

\bibitem[{Robertson {et~al.}(2013)Robertson, Furlanetto, Schneider, Charlot,
  Ellis, Stark, McLure, Dunlop, Koekemoer, Schenker, Ouchi, Ono, Curtis-Lake,
  Rogers, Bowler, \& Cirasuolo}]{Robertson2013}
Robertson, B.~E. {et~al.} 2013, \apj, 768, 71

\bibitem[{{Rollinde} {et~al.}(2009){Rollinde}, {Vangioni}, {Maurin}, {Olive},
  {Daigne}, {Silk}, \& {Vincent}}]{Rollinde2009}
{Rollinde}, E., {Vangioni}, E., {Maurin}, D., {Olive}, K.~A., {Daigne}, F.,
  {Silk}, J., \& {Vincent}, F.~H. 2009, \mnras, 398, 1782

\bibitem[{{Salvadori} {et~al.}(2007){Salvadori}, {Schneider}, \&
  {Ferrara}}]{Salvadori2007}
{Salvadori}, S., {Schneider}, R., \& {Ferrara}, A. 2007, \mnras, 381, 647

\bibitem[{{Schaerer}(2002)}]{Schaerer2002}
{Schaerer}, D. 2002, \aap, 382, 28

\bibitem[{{Seager} {et~al.}(1999){Seager}, {Sasselov}, \& {Scott}}]{Seager1999}
{Seager}, S., {Sasselov}, D.~D., \& {Scott}, D. 1999, \apjl, 523, L1

\bibitem[{{Seager} {et~al.}(2000){Seager}, {Sasselov}, \& {Scott}}]{Seager2000}
---. 2000, \apjs, 128, 407

\bibitem[{{Shull} \& {van Steenberg}(1985)}]{Shull1985}
{Shull}, J.~M. \& {van Steenberg}, M.~E. 1985, \apj, 298, 268

\bibitem[{{Sigurdson} \& {Furlanetto}(2006)}]{Sigurdson2006}
{Sigurdson}, K. \& {Furlanetto}, S.~R. 2006, Physical Review Letters, 97,
  091301

\bibitem[{Spergel {et~al.}(2003)Spergel, Verde, Peiris, Komatsu, Nolta,
  Bennett, Halpern, Hinshaw, Jarosik, \& Kogut}]{Spergel2003}
Spergel, D.~N. {et~al.} 2003, \apjs, 148, 175

\bibitem[{Stern(1999)}]{Stern1999}
Stern, S.~A. 1999, Reviews of Geophysics, 37, 453

\bibitem[{{Tanaka} {et~al.}(2012){Tanaka}, {Perna}, \& {Haiman}}]{Tanaka2012}
{Tanaka}, T., {Perna}, R., \& {Haiman}, Z. 2012, \mnras, 425, 2974

\bibitem[{Tegmark {et~al.}(1997)Tegmark, Silk, Rees, Blanchard, Abel, \&
  Palla}]{Tegmark1997}
Tegmark, M., Silk, J., Rees, M.~J., Blanchard, A., Abel, T., \& Palla, F. 1997,
  \apj, 474, 1

\bibitem[{Thomas \& Zaroubi(2008)}]{Thomas2008}
Thomas, R.~M. \& Zaroubi, S. 2008, \mnras, 384, 1080

\bibitem[{{Tseliakhovich} \& {Hirata}(2010)}]{Tseliakhovich2010}
{Tseliakhovich}, D. \& {Hirata}, C. 2010, \prd, 82, 083520

\bibitem[{Tumlinson \& Shull(2000)}]{Tumlinson2000}
Tumlinson, J. \& Shull, J.~M. 2000, \apj, 528, L65

\bibitem[{{van Haarlem} {et~al.}(2013){van Haarlem}, {Wise}, {Gunst}, {Heald},
  {McKean}, {Hessels}, {de Bruyn}, {Nijboer}, {Swinbank}, {Fallows},
  {Brentjens}, {Nelles}, {Beck}, {Falcke}, {Fender}, {H{\"o}randel},
  {Koopmans}, {Mann}, {Miley}, {R{\"o}ttgering}, {Stappers}, {Wijers},
  {Zaroubi}, {van den Akker}, {Alexov}, {Anderson}, {Anderson}, {van Ardenne},
  {Arts}, {Asgekar}, {Avruch}, {Batejat}, {B{\"a}hren}, {Bell}, {Bell}, {van
  Bemmel}, {Bennema}, {Bentum}, {Bernardi}, {Best}, {B{\^i}rzan}, {Bonafede},
  {Boonstra}, {Braun}, {Bregman}, {Breitling}, {van de Brink}, {Broderick},
  {Broekema}, {Brouw}, {Br{\"u}ggen}, {Butcher}, {van Cappellen}, {Ciardi},
  {Coenen}, {Conway}, {Coolen}, {Corstanje}, {Damstra}, {Davies}, {Deller},
  {Dettmar}, {van Diepen}, {Dijkstra}, {Donker}, {Doorduin}, {Dromer}, {Drost},
  {van Duin}, {Eisl{\"o}ffel}, {van Enst}, {Ferrari}, {Frieswijk}, {Gankema},
  {Garrett}, {de Gasperin}, {Gerbers}, {de Geus}, {Grie{\ss}meier}, {Grit},
  {Gruppen}, {Hamaker}, {Hassall}, {Hoeft}, {Holties}, {Horneffer}, {van der
  Horst}, {van Houwelingen}, {Huijgen}, {Iacobelli}, {Intema}, {Jackson},
  {Jelic}, {de Jong}, {Juette}, {Kant}, {Karastergiou}, {Koers}, {Kollen},
  {Kondratiev}, {Kooistra}, {Koopman}, {Koster}, {Kuniyoshi}, {Kramer},
  {Kuper}, {Lambropoulos}, {Law}, {van Leeuwen}, {Lemaitre}, {Loose}, {Maat},
  {Macario}, {Markoff}, {Masters}, {McKay-Bukowski}, {Meijering}, {Meulman},
  {Mevius}, {Middelberg}, {Millenaar}, {Miller-Jones}, {Mohan}, {Mol},
  {Morawietz}, {Morganti}, {Mulcahy}, {Mulder}, {Munk}, {Nieuwenhuis}, {van
  Nieuwpoort}, {Noordam}, {Norden}, {Noutsos}, {Offringa}, {Olofsson}, {Omar},
  {Orr{\'u}}, {Overeem}, {Paas}, {Pandey-Pommier}, {Pandey}, {Pizzo},
  {Polatidis}, {Rafferty}, {Rawlings}, {Reich}, {de Reijer}, {Reitsma},
  {Renting}, {Riemers}, {Rol}, {Romein}, {Roosjen}, {Ruiter}, {Scaife}, {van
  der Schaaf}, {Scheers}, {Schellart}, {Schoenmakers}, {Schoonderbeek},
  {Serylak}, {Shulevski}, {Sluman}, {Smirnov}, {Sobey}, {Spreeuw}, {Steinmetz},
  {Sterks}, {Stiepel}, {Stuurwold}, {Tagger}, {Tang}, {Tasse}, {Thomas},
  {Thoudam}, {Toribio}, {van der Tol}, {Usov}, {van Veelen}, {van der Veen},
  {ter Veen}, {Verbiest}, {Vermeulen}, {Vermaas}, {Vocks}, {Vogt}, {de Vos},
  {van der Wal}, {van Weeren}, {Weggemans}, {Weltevrede}, {White}, {Wijnholds},
  {Wilhelmsson}, {Wucknitz}, {Yatawatta}, {Zarka}, {Zensus}, \& {van
  Zwieten}}]{vanHaarlem2013}
{van Haarlem}, M.~P. {et~al.} 2013, preprint (astroph/13053550)

\bibitem[{Vedantham {et~al.}(2013)Vedantham, Koopmans, de~Bruyn, Wijnholds,
  Ciardi, \& Brentjens}]{Vedantham2013}
Vedantham, H.~K., Koopmans, L. V.~E., de~Bruyn, A.~G., Wijnholds, S.~J.,
  Ciardi, B., \& Brentjens, M.~A. 2013, preprint (astroph/13062172)

\bibitem[{Venkatesan \& Benson(2011)}]{Venkatesan2011}
Venkatesan, A. \& Benson, A. 2011, \mnras, 417, 2264

\bibitem[{{Whalen} {et~al.}(2013{\natexlab{a}}){Whalen}, {Even}, {Lovekin},
  {Fryer}, {Stiavelli}, {Roming}, {Cooke}, {Pritchard}, {Holz}, \&
  {Knight}}]{Whalen2013SNIIn}
{Whalen}, D.~J. {et~al.} 2013{\natexlab{a}}, \apj, 768, 195

\bibitem[{{Whalen} {et~al.}(2013{\natexlab{b}}){Whalen}, {Fryer}, {Holz},
  {Heger}, {Woosley}, {Stiavelli}, {Even}, \& {Frey}}]{Whalen2013PISN}
{Whalen}, D.~J., {Fryer}, C.~L., {Holz}, D.~E., {Heger}, A., {Woosley}, S.~E.,
  {Stiavelli}, M., {Even}, W., \& {Frey}, L.~H. 2013{\natexlab{b}}, \apjl, 762,
  L6

\bibitem[{{Wouthuysen}(1952)}]{Wouthuysen1952}
{Wouthuysen}, S.~A. 1952, \aj, 57, 31

\bibitem[{Zackrisson {et~al.}(2011)Zackrisson, Rydberg, Schaerer, {\"O}stlin,
  \& Tuli}]{Zackrisson2011}
Zackrisson, E., Rydberg, C.-E., Schaerer, D., {\"O}stlin, G., \& Tuli, M. 2011,
  \apj, 740, 13

\bibitem[{Zackrisson {et~al.}(2012)Zackrisson, Zitrin, Trenti, Rydberg, Guaita,
  Schaerer, Broadhurst, {\"O}stlin, \& Str{\"o}m}]{Zackrisson2012}
Zackrisson, E. {et~al.} 2012, \mnras, 427, 2212

\bibitem[{{Zahn} {et~al.}(2011){Zahn}, {Mesinger}, {McQuinn}, {Trac}, {Cen}, \&
  {Hernquist}}]{Zahn2011}
{Zahn}, O., {Mesinger}, A., {McQuinn}, M., {Trac}, H., {Cen}, R., \&
  {Hernquist}, L.~E. 2011, \mnras, 414, 727

\bibitem[{{Zahn} {et~al.}(2012){Zahn}, {Reichardt}, {Shaw}, {Lidz}, {Aird},
  {Benson}, {Bleem}, {Carlstrom}, {Chang}, {Cho}, {Crawford}, {Crites}, {de
  Haan}, {Dobbs}, {Dor{\'e}}, {Dudley}, {George}, {Halverson}, {Holder},
  {Holzapfel}, {Hoover}, {Hou}, {Hrubes}, {Joy}, {Keisler}, {Knox}, {Lee},
  {Leitch}, {Lueker}, {Luong-Van}, {McMahon}, {Mehl}, {Meyer}, {Millea},
  {Mohr}, {Montroy}, {Natoli}, {Padin}, {Plagge}, {Pryke}, {Ruhl}, {Schaffer},
  {Shirokoff}, {Spieler}, {Staniszewski}, {Stark}, {Story}, {van Engelen},
  {Vanderlinde}, {Vieira}, \& {Williamson}}]{Zahn2012}
{Zahn}, O. {et~al.} 2012, \apj, 756, 65

\bibitem[{{Zheng} {et~al.}(2012){Zheng}, {Postman}, {Zitrin}, {Moustakas},
  {Shu}, {Jouvel}, {H{\o}st}, {Molino}, {Bradley}, {Coe}, {Moustakas},
  {Carrasco}, {Ford}, {Ben{\'{\i}}tez}, {Lauer}, {Seitz}, {Bouwens},
  {Koekemoer}, {Medezinski}, {Bartelmann}, {Broadhurst}, {Donahue}, {Grillo},
  {Infante}, {Jha}, {Kelson}, {Lahav}, {Lemze}, {Melchior}, {Meneghetti},
  {Merten}, {Nonino}, {Ogaz}, {Rosati}, {Umetsu}, \& {van der Wel}}]{Zheng2012}
{Zheng}, W. {et~al.} 2012, \nat, 489, 406

\bibitem[{Zygelman(2005)}]{Zygelman2005}
Zygelman, B. 2005, \apj, 622, 1356

\end{thebibliography}
\bibliographystyle{apj_w_etal}

\end{document}